\documentclass[showpacs,nofootinbib]{revtex4}
\usepackage{graphicx}
\usepackage{latexsym}

\def\be{\begin{equation}}
\def\ee{\end{equation}}
\def\bea{\begin{eqnarray}}
\def\eea{\end{eqnarray}}
\newcommand{\DE}{{\mathrm{DE}}}
\newcommand{\DM}{{\mathrm{DM}}}

\newcommand{\eff}{{\mathrm{eff}}}
\newcommand{\sys}{{\mathrm{sys}}}
\newcommand{\maga}{{\mathrm{mag}}}
\newcommand{\sky}{{\mathrm{sky}}}
\newcommand{\maxa}{{\mathrm{max}}}
\newcommand{\FWHM}{{\mathrm{fwhm}}}

\begin{document}

\title{Probing for the Cosmological Parameters with PLANCK Measurement}

\author{Jun-Qing Xia${}^a$}
\email{xiajq@mail.ihep.ac.cn}

\author{Hong Li${}^b$}
\author{Gong-Bo Zhao${}^a$}
\author{Xinmin Zhang${}^a$}

\affiliation{${}^a$Institute of High Energy Physics, Chinese
Academy of Science, P.O. Box 918-4, Beijing 100049, P. R. China}

\affiliation{${}^b$Department of Astronomy, School of Physics,
Peking University, Beijing 100871, P. R. China}

\date{\today}

\begin{abstract}

In this paper we investigate the constraints on the cosmological
parameters, especially for the equation of state of dynamical dark
energy $w_{\DE}$, the inflationary parameters $n_s$, $\alpha_s$
and $r$, the total neutrino mass $\sum m_{\nu}$ and the curvature
of universe $\Omega_{K}$, using the simulated data of future
Planck measurement. Firstly we determine the cosmological
parameters with the current observations, including ESSENCE (192
sample), WMAP three year (WMAP3), Boomerang-2K2, CBI, VSA, ACBAR,
SDSS LRG and 2dFGRS, and then take the best-fit model as the
fiducial model in our following simulations. In the simulations we
pay particular attention to the effects of the dynamical dark
energy in the determination of the cosmological parameters. Due to
this reason, in order to make our constraints more robust, we have
added the simulated SNAP data in our simulations. Using the
present data, we find that the Quintom dark energy model is mildly
favored, while the $\Lambda$CDM model remains a good fit. In the
framework of dynamical dark energy model, the constraints on the
inflationary parameters, $\sum m_{\nu}$ and $\Omega_{K}$ become
weak, compared with the constraints in the $\Lambda$CDM model.
Intriguingly, we find that the inflationary models with a ``blue"
tilt, which are excluded about $2\sigma$ in the $\Lambda$CDM
model, are well within $2\sigma$ region with the presence of the
dynamics of dark energy. The upper limits of neutrino mass are
weakened by a factor of 2 $(95\%~C.L.)$, say, $\sum
m_{\nu}<1.59~eV$ and $\sum m_{\nu}<1.53~eV$ for two forms of
parametrization of the equation of state of dark energy. The flat
universe is a good fit to the current data, namely,
$|\Omega_{K}|<0.03~(95\%~C.L.)$. With the simulated Planck and
SNAP data, the dynamical dark energy model and the $\Lambda$CDM
model might be distinguished at 4$\sigma$ confidence level. And
the uncertainties of the inflationary parameters, $\sum m_{\nu}$
and $\Omega_{K}$ can be reduced significantly in the framework of
the dynamical dark energy model. We also constrain the rotation
angle $\Delta\alpha$, denoting the possible \emph{CPT} violation,
from the simulated Planck and CMBpol data and find that our
results are much more stringent than the current constraint and
will be used to verify the \emph{CPT} symmetry with a higher
precision.

\end{abstract}


\maketitle


\section{Introduction}
\label{Int}

With the accumulation of observational data, such as the
Supernovae Type-Ia (SN Ia) \cite{Miknaitis:2007jd,Davis:2007na},
Cosmic Microwave Background (CMB)
\cite{wmap3:2006:1,wmap3:2006:2,wmap3:2006:3,wmap3:2006:4}, Large
Scale Structure (LSS) \cite{Tegmark:2006az} and so forth, it is
possible for us to unveil, despite not conclusively for the time
being, some enigmas in cosmology, such as the nature of dark
energy (DE), the inflation, the total neutrino mass $\sum
m_{\nu}$, the curvature of our Universe $\Omega_{K}$ and even the
possible violation of \emph{CPT} conservation in Cosmology
\cite{Feng:2006dp}. Dark energy, the mysterious source driving the
present acceleration of our Universe, has been studied widely in
the literature since its first discovery in $1998$ \cite{SN98}.
However, the nature of DE, encoded in its equation of state (EoS)
parameter $w$, remains controversial. Being the simplest candidate
of DE and fitting the current data well, the Cosmological Constant
(CC), whose EoS remains $-1$, suffers from the severe theoretical
drawbacks such as the fine-tunning and coincidence problems
\cite{SW89}. To ameliorate such problems, the dynamical dark
energy models were proposed. For example, the Quintessence, whose
EoS evolves with the cosmic time and satisfies $w(z)>-1$
\cite{quint}, Phantom with $w(z)<-1$ \cite{phantom} and K-essence
with $w(z)>-1$ or $<-1$ \cite{kessence}. As addressed in
literature, recent observations mildly favor the DE models with
$w(z)$ crossing the cosmological constant boundary during the
evolution
\cite{quintom,Xia:2005ge,Xia:2006cr,Zhao:2006bt,Xia:2006wd,Zhao:2006qg,Xia:2007km,others}.
Unfortunately, the EoS of the above models cannot realize such
``crossing" behavior due to the ``No-Go" Theorem
\cite{Xia:2007km,Vikman:2004dc,Kunz:2006wc}. Quintom, whose EoS
can cross the cosmological constant boundary, is mildly favored by
the observations and has been investigating extensively since its
invention \cite{quintom,study4quintom}.

Our universe has experienced at least two different stages of
accelerated expansion. One is the current acceleration driven by
dark energy, the other is the inflation in the very early universe
\cite{inflation,Guth:1979bh}. The mechanics of inflation can
naturally explain the flatness, homogeneity and the isotropy of
our Universe. Inflation stretches the primordial density
fluctuations and seeds the presently observed large scale
structures and cosmic microwave background radiation. In $2006$,
the WMAP group claimed that the simple scale-invariant primordial
spectrum does not fit well to the Three-Year WMAP data
\cite{wmap3:2006:1}. Alternatively, the
Harrison-Zel'dovich-Peebles scale invariant (HZ) spectrum
($n_s=1,~r=0$) is disfavored about $2 \sim 3\sigma$. And the large
running of the scalar spectral index is still allowed
\cite{Easther:2006tv}.

The aforementioned key cosmological questions might be answered by
the virtue of the future high precision astronomical measurements.
Especially, the Planck mission of European Space Agency (ESA) will
determine the geometry and contents of our Universe by measuring
the CMB with unprecedented accuracy \cite{:2006uk}. Planck will
image the full sky with sensitivity of $\Delta
T/T\sim2\times10^{-6}$, angular resolution to $5'$ and frequency
coverage of $30-857$ GHz. The angular resolution of Planck
measurement is three times superior to the current WMAP
observation and the noise is lowered by an order of magnitude at
around $100$ GHz. These significant improvements permit much more
accurate measurements of the CMB power spectra, so that Planck has
the very power and unique new capabilities to constrain the
cosmological parameters. In Ref.\cite{:2006uk}, the Planck
collaboration has done some sensitivity studies of constraining
the cosmological parameters with simulated Planck data combined
with the future SNAP measurement. They investigate the dynamics of
inflation, neutrino mass, \emph{etc.} in the framework of the
$\Lambda$CDM model and find that with Planck one can get much more
stringent constraints on the cosmological parameters.

In our previous works \cite{Xia:2006cr,Xia:2006wd,Zhao:2006qg} we
addressed that the determination of the cosmological parameters,
such as $\sum m_{\nu}$ , $\Omega_{K}$ and the inflationary
parameters, is highly affected by the dynamics of dark energy
model due to the degeneracies among the EoS of DE and these
parameters. Furthermore, dark energy perturbation plays a crucial
role in the global fit \cite{Xia:2005ge,Zhao:2005vj}. Therefore,
it is much more fair and reliable to do the error forecasts of the
cosmological parameters in the framework of dynamical dark energy
model rather than assuming a constant $w$ of DE or the
$\Lambda$CDM model. In this paper, we study the constraints of
$\sum m_{\nu}$ , $\Omega_{K}$ as well as the inflationary
parameters in the framework of dynamical dark energy models. Using
the simulated Planck data, we make a global fit using MCMC method,
while paying particular attention of the dark energy perturbation
in the full parameter space of EoS of dark energy. We also stress
the role of Planck and CMBpol to detect the possible \emph{CPT}
violation. To obtain the fiducial value of parameters for
simulation, we firstly constrain these cosmological parameters
from the current observations and find the best-fit models.

Our paper is organized as follows: In Section II we describe the
method and the current observational datasets we used; In Section
III we present our method to do the futuristic simulations in
detail; Section IV contains our MCMC fitting results using the
current and future observations and the last section is our
conclusion and discussion.


\section{Method and Current Observations}
\label{Method}

In our studies, we have modified the publicly available Markov
Chain Monte Carlo package \emph{CosmoMC}\footnote{Available at:
http://cosmologist.info/.} \cite{CosmoMC} to include the dark
energy perturbation when the EoS of DE gets across the
cosmological constant boundary as we illustrate later. We assume
the purely adiabatic initial conditions. Our most general
parameter space is:
\begin{equation}
\label{parameter} {\bf P} \equiv (\omega_{b}, \omega_{c},
\Omega_k, \Theta_{s}, \tau, w_{0}, w_{1}, f_{\nu}, n_{s},
\log[10^{10}A_{s}], \alpha_s, r)~,
\end{equation}
where $\omega_{b}\equiv\Omega_{b}h^{2}$ and
$\omega_{c}\equiv\Omega_{c}h^{2}$ are the physical baryon and cold
dark matter densities relative to the critical density,
$\Omega_k=1-\Omega_m-\Omega_{\DE}$ is the spatial curvature,
$\Theta_{s}$ is the ratio (multiplied by 100) of the sound horizon
to the angular diameter distance at decoupling, $\tau$ is the
optical depth to re-ionization, $f_{\nu}$ is the dark matter
neutrino fraction at present, namely,
\begin{equation}
f_{\nu}\equiv\frac{\rho_{\nu}}{\rho_{\DM}}=\frac{\Sigma
m_{\nu}}{93.105~eV~\Omega_ch^2}~,
\end{equation}
$A_{s}$ is defined as the amplitude of the primordial spectrum. We
parameterize the primordial power spectrum in form of
\begin{equation}
\label{ns} n_s(k)=n_s(k_{s0}) + \alpha_{s} \ln \left(
\frac{k}{k_{s0}}\right)~,
\end{equation}
where $k_{s0}$ is a pivot scale which is arbitrary in principle,
here we set $k_{s0}=0.05$Mpc$^{-1}$, and $\alpha_{s}$ is a
constant characterizing the ``running" $dn_s/d\ln k$ of the scalar
spectral index. $r$ is the tensor to scalar ratio of the
primordial spectrum. The scalar spectral index $n_s$ is related to
the primordial scalar power spectrum ${\cal P}_{\chi}(k)$ by
definition:
\begin{equation}
\label{nsdef} n_s(k) \equiv \frac{d{\cal P}_{\chi}(k)}{d \ln k}
+1~.
\end{equation}
Correspondingly, ${\cal P}_{\chi}(k)$ is now parameterized as
\cite{paraPk}:
\begin{equation}
\label{spectrum} \ln {\cal P}_{\chi}(k)=\ln A_{s} +
(n_{s}(k_{s0})-1)\ln \left(
\frac{k}{k_{s0}}\right)+\frac{\alpha_{s}}{2}\left(\ln
\left(\frac{k}{k_{s0}}\right)\right)^{2}~.
\end{equation}

\begin{figure}[htbp]
\begin{center}
\includegraphics[scale=0.7]{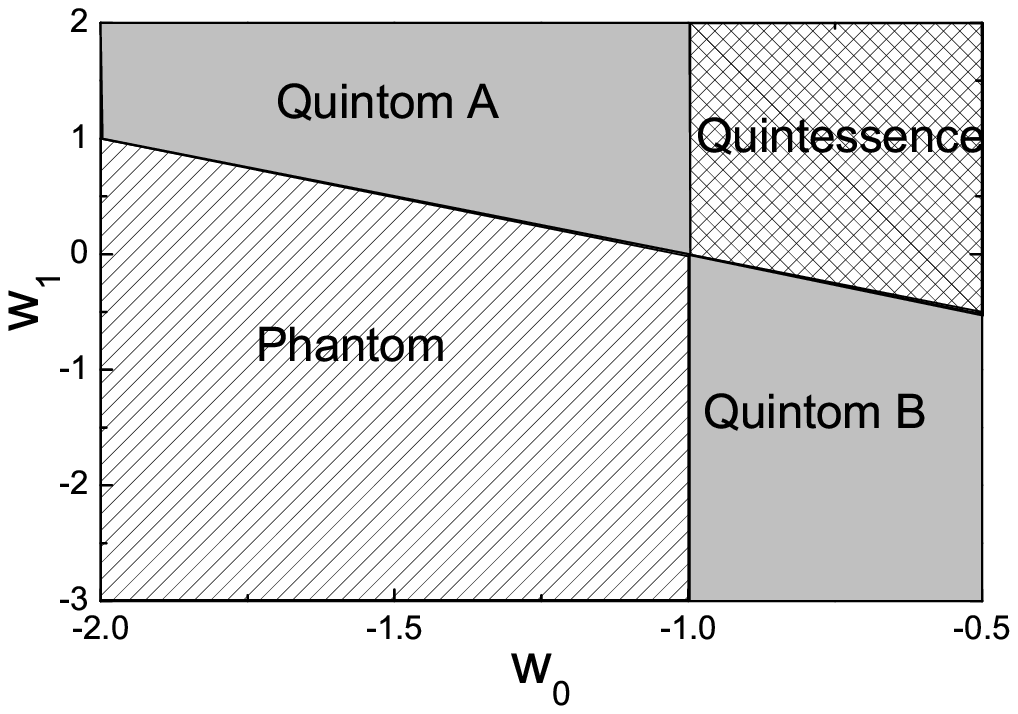}
\includegraphics[scale=0.7]{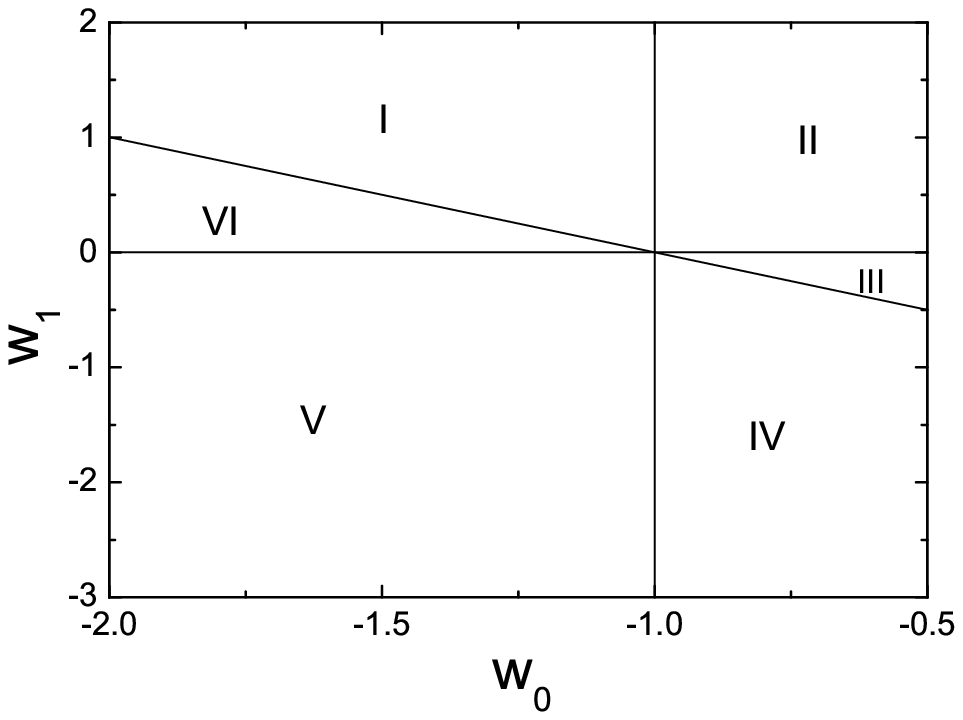}
\caption{Left panel: The parameter space is divide into four parts
to distinguish different dark energy models; Right panel: The
parameter space is divide into six parts including the future
behavior of the EoS of dark energy. See text for
details.\label{fig0}}
\end{center}
\end{figure}

For dark energy, we choose the commonly used parametrization of the
DE equation of state as \cite{Linderpara}:
\begin{equation}
\label{EOS1} w_\DE(a) = w_{0} + w_{1}(1-a)~,
\end{equation}
where $a=1/(1+z)$ is the scale factor and $w_{1}=-dw/da$
characterizes the ``running" of the equation of state. In left
panel of Fig.\ref{fig0}, we divide the ($w_0$,~$w_1$) panel into
four blocks by lines $w_0=-1$ and $w_0+w_1=-1$ as illustrated. In
the upper right and lower left parts, $w(z)$ is greater or smaller
than $-1$ corresponding to the Quintessence and Phantom models
respectively. In the other two parts, $w(z)$ can cross the
cosmological constant boundary during the evolution which can be
realized by the Quintom model. The models of Quintom A crosses
$-1$ from upside down while Quintom B crosses from the other
direction during the evolution. And the intersecting point denotes
the $\Lambda$CDM model. However, if one takes the futuristic
evolution of EoS into consideration, parts of the region occupied
by the Quintessence and Phantom will be replaced by Quintom. More
explicitly, in the right panel of Fig.\ref{fig0}, we redivide the
parameter space into six parts by the lines $w_0=-1$, $w_0+w_1=-1$
and $w_1=0$. Part III is for the Quintessence-like models, namely,
the equation of state remains greater than $-1$ regardless of the
cosmic time, say, $w>-1$ for past, present and future.
Correspondingly, part VI is for the Phantom-like models. Part
I,II,V and IV are all for the Quintom-like models. For the models
lie within part I and IV, their equation of state has crossed over
$-1$ till now while the EoS of the DE models in part II and V will
cross $-1$ in future.

Moreover, we also consider another phenomenological
parametrization, namely, oscillating Quintom whose EoS oscillates
with time and is allowed to cross the cosmological constant
boundary:
\begin{equation}
\label{EOS2} w_\DE(a) = w_{0} + w_{1}\sin(w_2\ln(a))~.
\end{equation}
This oscillating behavior in EoS can lead to the oscillation on
the Hubble diagram \cite{Xia:2006rr} or a recurrent universe which
can unify the early inflation and the current acceleration of our
Universe \cite{Feng:2004ff}. In
Refs.\cite{Barenboim:2004kz,Xia:2004rw,Xia:2006rr,Zhao:2006qg}
some preliminary studies have been presented on this kind of dark
energy model. From the latest SNIa paper \cite{Riess:2006fw}, one
can find some hints of oscillating behavior in their Fig.$10$
where they used a polynomial fitting. Our sine function has the
advantage of preserving the oscillating feature of the dark energy
EoS at high redshift measured by the CMB data. For simplicity and
focusing on the study at lower redshift, we set $w_2$ to be
$3\pi/2$ in order to allow the EoS to evolve more than one period
within the redshift range of $z=0$ to $z=2$ where the SNIa data
are most robust. We label the above two dark energy
parameterizations (\ref{EOS1}) and (\ref{EOS2}) as Para I and Para
II respectively throughout this paper.

When using the MCMC global fitting strategy to constrain the
cosmological parameters, it is crucial to include Dark Energy
perturbation. The conservation law of energy reads:
\begin{equation}
\label{conserve} T^{\mu\nu}{}_{;\mu}=0~,
\end{equation}
where $T^{\mu\nu}$ is the energy-momentum tensor of dark energy
and ``;" denotes the covariant differentiation. Working in the
conformal Newtonian gauge and normally setting the anisotropic
stress perturbation of dark energy to be zero, one can derive the
perturbation equations of dark energy as follows \cite{ma}:
\begin{eqnarray}
\delta'&=&-(1+w)(\theta-3\Phi')
-3\mathcal{H}(\hat{c}_{s}^2-w)\delta-3\mathcal{H}(w'+3\mathcal{H}(1+w)(\hat{c}_{s}^2-w))\frac{\theta}{k^2}~, \label{dotdelta}\\
\theta'&=&-\mathcal{H}(1-3\hat{c}_{s}^2)\theta+k^{2}(\frac{\hat{c}_{s}^2\delta}{{1+w}}+
\Psi)~. \label{dottheta}
\end{eqnarray}
However one cannot handle the Dark Energy perturbation when the
parameterized EoS crosses $-1$ based on the Quintessence, Phantom,
K-essence and other non-crossing dark energy models. By the virtue
of Quintom dark energy model, the perturbation at the crossing
points is continuous, thus we introduce a small positive constant
$\epsilon$ to divide the full range of the allowed value of $w$
into three parts: 1) $ w > -1 + \epsilon$; 2) $-1 + \epsilon \geq
w  \geq-1 - \epsilon$; and 3) $w < -1 -\epsilon $. Neglecting the
entropy perturbation contributions, for the regions 1) and 3) the
equation of state does not get across $-1$ and the perturbations
are well defined by solving Eqs.(\ref{dotdelta},\ref{dottheta}).
For the case 2), the density perturbation $\delta$ and velocity
perturbation $\theta$, and the derivatives of $\delta$ and
$\theta$ are finite and continuous for the realistic Quintom dark
energy models. However for the perturbations of the parameterized
EoS there is clearly a divergence. In our study for such a regime,
we match the perturbation in region 2) to the regions 1) and 3) at
the boundary and set:
\begin{equation}\label{dotx}
\delta'=0 ~~,~~\theta'=0~~.
\end{equation}
In our numerical calculations we have limited the range to be $|
\epsilon |<10^{-5}$ and we find our method is a very good
approximation to the multi-field Quintom DE model. In our
calculation the initial condition we choose is the adiabatic
perturbations of dark energy, while the isocurvature perturbation
of dark energy can be safely neglected \cite{Zhao:2005vj}. For
more details of this method we refer the readers to our previous
companion papers \cite{Zhao:2005vj,Xia:2005ge}.

In our calculations we have taken the total likelihood to be the
products of the separate likelihoods (${\bf \cal{L}}_i$) of CMB,
LSS and SNIa. In other words defining $\chi_{L,i}^2 \equiv -2 \log
{\bf \cal{L}}_i$, we get
\begin{equation}
\chi^2_{L,total} = \chi^2_{L,CMB} + \chi^2_{L,LSS} +
\chi^2_{L,SNIa}~.\label{chi2}
\end{equation}
If the likelihood function is Gaussian, $\chi^2_{L}$ coincides
with the usual definition of $\chi^2$ up to an additive constant
corresponding to the logarithm of the normalization factor of
${\cal L}$. In the computation of CMB we have included the
three-year WMAP (WMAP3) Temperature-Temperature (TT) and
Temperature-Polarization (TE) power spectrum with the routine for
computing the likelihood supplied by the WMAP team
\cite{wmap3:2006:1,wmap3:2006:2,wmap3:2006:3,wmap3:2006:4} as well
as the smaller scale experiments, including Boomerang-2K2
\cite{MacTavish:2005yk}, CBI \cite{Readhead:2004gy}, VSA
\cite{Dickinson:2004yr} and ACBAR \cite{Kuo:2002ua}. For the Large
Scale Structure information, we have used the Sloan Digital Sky
Survey (SDSS) luminous red galaxy (LRG) sample
\cite{Tegmark:2006az} and 2dFGRS \cite{Cole:2005sx}. To be
conservative but more robust, in the fittings to the SDSS LRG
sample we have used the first $15$ bins only, which are supposed
to be well within the linear regime. In the calculation of the
likelihood from SNIa we have marginalized over the nuisance
parameter \cite{DiPietro:2002cz}. The supernova data we use are
the recently released ESSENCE (192 sample) data
\cite{Miknaitis:2007jd,Davis:2007na}. Furthermore, we make use of
the Hubble Space Telescope (HST) measurement of the Hubble
parameter $H_{0}\equiv 100$h~km~s$^{-1}$~Mpc$^{-1}$ \cite{Hubble}
by multiplying the likelihood by a Gaussian likelihood function
centered around $h=0.72$ and with a standard deviation
$\sigma=0.08$. We also impose a weak Gaussian prior on the baryon
density $\Omega_{b}h^{2}=0.022\pm0.002$ (1 $\sigma$) from the Big
Bang Nucleosynthesis \cite{BBN}. Simultaneously we will also use a
cosmic age tophat prior as 10 Gyr $< t_0 <$ 20 Gyr.

For each regular calculation, we run 8 independent chains
comprising of $150,000-300,000$ chain elements and spend thousands
of CPU hours to calculate on a supercomputer. The average
acceptance rate is about $40\%$. We test the convergence of the
chains by Gelman and Rubin criteria\cite{R-1} and find $R-1$ is of
order $0.01$ which is more conservative than the recommended value
$R-1<0.1$.


\section{Future Measurements}
\label{Future}

When considering the constraints on the cosmological parameters
from the simulated data of future CMB (PLANCK\footnote{Available
at http://sci.esa.int/science-e/www/area/index.cfm?fareaid=17/.}
and CMBpol\footnote{Available at
http://universe.gsfc.nasa.gov/program/inflation.html/.})
measurements, the fiducial models are obtained by maximizing the
likelihood (the best-fit model) using the current observations.


Firstly we derive the likelihood function for a CMB experiment as
given by Ref.\cite{Easther:2004vq,Perotto:2006rj}. Assuming the
CMB multipoles are Gaussian distributed, one can obtain the
likelihood function as follows:
\begin{equation}
\mathcal{L}\propto\prod_{lm}\frac{\exp \left[
-\frac{1}{2}D^{\dag}_{lm}C^{-1}D_{lm}\right]}{\sqrt{\det
C}}~,\label{CMB1}
\end{equation}
where $D_{lm}=\left[ a^{T}_{lm}, a^{E}_{lm}, a^{B}_{lm} \right]$
is the data vector of spherical harmonic coefficients which are
contributed from the CMB signal $s_{lm}$ and the experimental
noise $n_{lm}$: $a^{X}_{lm}=s^{X}_{lm}+n^{X}_{lm}$, and $C$ is the
theoretical data covariance matrix generally given by:
\begin{equation}
C=\left( \begin{array}{ccc}
    \bar{C}^{TT}_{l}&\bar{C}^{TE}_{l}&\bar{C}^{TB}_{l} \\
    \bar{C}^{TE}_{l}&\bar{C}^{EE}_{l}&\bar{C}^{EB}_{l} \\
    \bar{C}^{TB}_{l}&\bar{C}^{EB}_{l}&\bar{C}^{BB}_{l} \\
\end{array} \right)=\left( \begin{array}{ccc}
    C^{TT}_{l}+N^{TT}_{l}&C^{TE}_{l}&C^{TB}_{l} \\
    C^{TE}_{l}&C^{EE}_{l}+N^{EE}_{l}&C^{EB}_{l} \\
    C^{TB}_{l}&C^{EB}_{l}&C^{BB}_{l}+N^{BB}_{l} \\
\end{array} \right)~.\label{CMB2}
\end{equation}
In this covariance matrix, $C^{XX'}_{l}$ denotes the theoretical
power spectra and $N^{XX'}_{l}$ is the noise power spectra which can
be approximated as:
\begin{equation}
N^{XX'}_{l}\equiv\langle n^{X\dag}_{lm}n^{X'}_{lm}
\rangle=\delta_{XX'}\theta^{2}_{\FWHM}\Delta^{2}_{X}\exp \left[
l(l+1)\frac{\theta^{2}_{\FWHM}}{8\ln 2}\right]~,\label{CMB3}
\end{equation}
where $\theta_{\FWHM}$ is the full width at half maximum of the
Gaussian beam, and $\Delta_{X}$ is the root mean square of the
instrumental noise. Non-diagonal noise terms are expected to be
zero since the noise contributions from different maps are
uncorrelated. Due to the global isotropy, the terms $C^{TB}_{l}$
and $C^{EB}_{l}$ are always set to be zero. In our calculations we
also assume them to be zero except for studying the possible
\emph{CPT} violation in the later section \ref{Angle}.

On the other hand, we can estimate the power spectra from the data
as follows:
\begin{equation}
\hat{C}^{XY}_{l}=\sum_{m}\frac{|a^{X\dag}_{lm}a^{Y}_{lm}|}{2l+1}~.\label{CMB4}
\end{equation}
So we can obtain the effective $\chi^2_{\eff}$:
\begin{equation}
\chi^2_{\eff}\equiv-2\ln
\mathcal{L}=\sum_{l}(2l+1)f_{\sky}\left(\frac{A}{|\bar{C}|}+\ln\frac{|\bar{C}|}{|\hat{C}|}+3\right)~,\label{CMB9}
\end{equation}
where $f_{\sky}$ denotes the observed fraction of the sky in the
real experiments, $A$ is defined as:
\begin{eqnarray}
A &=& \hat{C}^{TT}_{l}(\bar{C}^{EE}_{l}\bar{C}^{BB}_{l}-(\bar{C}^{EB}_{l})^2)+\hat{C}^{TE}_{l}(\bar{C}^{TB}_{l}\bar{C}^{EB}_{l}-\bar{C}^{TE}_{l}\bar{C}^{BB}_{l})+\hat{C}^{TB}_{l}(\bar{C}^{TE}_{l}\bar{C}^{EB}_{l}-\bar{C}^{TB}_{l}\bar{C}^{EE}_{l}) \nonumber\\
  &+& \hat{C}^{TE}_{l}(\bar{C}^{TB}_{l}\bar{C}^{EB}_{l}-\bar{C}^{TE}_{l}\bar{C}^{BB}_{l})+\hat{C}^{EE}_{l}(\bar{C}^{TT}_{l}\bar{C}^{BB}_{l}-(\bar{C}^{TB}_{l})^2)+\hat{C}^{EB}_{l}(\bar{C}^{TE}_{l}\bar{C}^{TB}_{l}-\bar{C}^{TT}_{l}\bar{C}^{EB}_{l}) \nonumber \\
  &+& \hat{C}^{TB}_{l}(\bar{C}^{TE}_{l}\bar{C}^{EB}_{l}-\bar{C}^{EE}_{l}\bar{C}^{TB}_{l})+\hat{C}^{EB}_{l}(\bar{C}^{TE}_{l}\bar{C}^{TB}_{l}-\bar{C}^{TT}_{l}\bar{C}^{EB}_{l})+\hat{C}^{BB}_{l}(\bar{C}^{TT}_{l}\bar{C}^{EE}_{l}-(\bar{C}^{TE}_{l})^2)~,\label{CMB6}
\end{eqnarray}
and $|\bar{C}|$ and $|\hat{C}|$ denote the determinants of the
theoretical and observed data covariance matrices respectively,
\begin{eqnarray}
|\bar{C}|&=&\bar{C}^{TT}_{l}\bar{C}^{EE}_{l}\bar{C}^{BB}_{l}+2\bar{C}^{TE}_{l}\bar{C}^{TB}_{l}\bar{C}^{EB}_{l}
           -\bar{C}^{TT}_{l}(\bar{C}^{EB}_{l})^2-\bar{C}^{EE}_{l}(\bar{C}^{TB}_{l})^2-\bar{C}^{BB}_{l}(\bar{C}^{TE}_{l})^2~,\label{CMB7}\\
|\hat{C}|&=&\hat{C}^{TT}_{l}\hat{C}^{EE}_{l}\hat{C}^{BB}_{l}+2\hat{C}^{TE}_{l}\hat{C}^{TB}_{l}\hat{C}^{EB}_{l}
           -\hat{C}^{TT}_{l}(\hat{C}^{EB}_{l})^2-\hat{C}^{EE}_{l}(\hat{C}^{TB}_{l})^2-\hat{C}^{BB}_{l}(\hat{C}^{TE}_{l})^2~.\label{CMB8}
\end{eqnarray}
The likelihood has been normalized with respect to the maximum
likelihood $\chi^2_{\eff}=0$, where
$\bar{C}^{XY}_{l}=\hat{C}^{XY}_{l}$. If we set the ${C}^{TB}_{l}$
and ${C}^{EB}_{l}$ to be zero, the likelihood function will be
reduced to the Eq.($17$) of Ref.\cite{Easther:2004vq}.
Furthermore, we can obtain the Eq.($9$) of Ref.\cite{Xia:2006wd}
if we ignore the tensor information.

In some of our simulations we also consider the gravitational
lensing effect on the CMB power spectrum. The lensed Stokes
parameters $I$, $Q$ and $U$ which specify the intensity and linear
polarization of observed CMB are related to the unlensed Stokes
parameters at the last scattering surface (denoted with a tilde)
by $X(\textbf{n})=\tilde X(\textbf{n}')=\tilde X(\textbf{n}+\delta
\textbf{n})$, where $X$ denotes $I$, $Q$ or $U$ and $\delta
\textbf{n}$ is the angular excursion of the photon as it
propagates from the last scattering surface until the present.
This deflection angle, $\delta \textbf{n}$, is given by the
gradient of the lensing potential
$\bigtriangledown\phi(\textbf{n})$,
\begin{equation}
\phi(\textbf{n})=2\int dr\frac{r-r_s}{rr_s}\Psi(r\hat \textbf{n},
r)~\label{lensphi},
\end{equation}
where $r$ is the comoving distance along the line of sight, $s$
denotes the CMB last scattering surface, and $\Psi$ is the three
dimensional gravitational potential
\cite{Zaldarriaga:1998ar,Lewis:2006fu}.

The important feature is that the gravitational lensing can mix $E$
and $B$ modes \cite{Zaldarriaga:1998ar}. If we assume that there is
only unlensed $E$ type polarization and the unlensed $\tilde
C_l^{BB}=0$ in the last scattering surface, the gravitational
lensing will generate $B$ type polarization in the observed field,
$C_l^{BB}\neq0$. The information from the gravitational lensing is
added through the power spectrum for the lensing potential
$C_l^{\phi\phi}$ and the correlation to the temperature
$C_l^{T\phi}$:
\begin{equation}
\langle
a^{\phi\dag}_{lm}a^{\phi}_{l'm'}\rangle=(C_l^{\phi\phi}+N_l^{\phi\phi})\delta_{ll'}\delta_{mm'}~~,~~
\langle
a^{T\dag}_{lm}a^{\phi}_{l'm'}\rangle=(C_l^{T\phi}+N_l^{T\phi})\delta_{ll'}\delta_{mm'}~,
\end{equation}
which can be computed numerically in the linear theory using
CAMB\footnote{Available at http://camb.info/.}
\cite{Lewis:1999bs}. In our analysis we use the unlensed power
spectra, $\tilde C_l^{TT}$, $\tilde C_l^{TE}$, $\tilde C_l^{EE}$,
and $C_l^{\phi\phi}$, $C_l^{T\phi}$. We do not use the lensed
power spectra to avoid the complication of the correlation in
their errors between different $l$ values and with the error in
$C_l^{\phi\phi}$ and $C_l^{T\phi}$ \cite{Hu:2001fb,Smith:2006nk}.
For errors on $C_l^{\phi\phi}$ we follow the Ref.\cite{Hu:2001kj}.
We use the publicly available code\footnote{Available at
http://lappweb.in2p3.fr/$\sim$perotto/FUTURCMB/home.html/.}
\cite{Perotto:2006rj} to simulate the mock CMB power spectra of
our fiducial models. In Table I we list the assumed experimental
specifications of the future Planck and CMBpol measurements and
neglect the foreground contamination.

\begin{table}
TABLE I. Assumed experimental specifications. We use the CMB power
spectra only at $l\leq2500$. The noise parameters $\Delta_T$ and
$\Delta_P$ are given in units of $\mu$K-arcmin.
\begin{center}
\begin{tabular}{lcccccc}
\hline \hline

~Experiment~ & ~$f_\sky$~ & ~$l_\maxa$~ & ~(GHz)~ &
~$\theta_\FWHM$~ & ~$\Delta_T$~ & ~$\Delta_P$~ \\

\hline

~PLANCK & 0.65 & 2500 & 100 & 9.5' & 6.8 & 10.9 \\
        &      &      & 143 & 7.1' & 6.0 & 11.4 \\
        &      &      & 217 & 5.0' & 13.1 & 26.7 \\
~CMBpol & 0.65 & 2500 & 217 & 3.0' & 1.0 & 1.4 \\

\hline \hline
\end{tabular}
\end{center}
\end{table}

To make our constraints more robust, we add the simulated SNAP
data to do all the simulations throughout this paper. The
projected satellite SNAP\footnote{SNAP is one of the several
candidates emission concepts for the Joint Dark Energy Mission
(JDEM). Available at http://snap.lbl.gov/.}
(Supernova/Acceleration Probe) would be a space-based telescope
with a one square degree field of view with $10^9$ pixels. It aims
to increase the discovery rate for SNIa to about $2000$ per year.
The simulated SNIa data distribution is taken from
Refs.\cite{kim,Li:2005zd}. As for the error, we follow the
Ref.\cite{kim} which takes the magnitude dispersion $0.15$ and the
systematic error $\sigma_{\sys}=0.02\times z/1.7$, and the whole
error for each data is as follows:
\begin{equation}
\sigma_{\maga}(z_i)=\sqrt{\sigma^2_{\sys}(z_i)+\frac{0.15^2}{n_i}}~,\label{snap}
\end{equation}
where $n_i$ is the number of supernova in the $i'$th redshift bin.


\section{Results}

In this section we show our global fitting results of the
cosmological parameters and focus on the dark energy parameters,
inflationary parameters, space-time curvature, total neutrino mass
and the rotation angle denoting the possible \emph{CPT} violation
respectively.


\subsection{Equation of State of Dark Energy}
\label{DEEoS}

\begin{table*}
TABLE II. Constraints on the EoS of dark energy and some
background parameters from the current observations and the future
simulations. Note that Para I and Para II represent $w_\DE(a) =
w_{0} + w_{1}(1-a)$ and $w_\DE(a) = w_{0} +
w_{1}\sin(3\pi/2\ln(a))$ respectively. For the current constraints
we have shown the mean values $1,2\sigma$ (Mean) and the best fit
results together. And we also list the standard deviation (SD) of
these parameters based on the future simulations.
\begin{center}

\begin{tabular}{|c|ccc|ccc|cc|}

\hline

&\multicolumn{3}{c|}{$\Lambda$CDM} &\multicolumn{3}{c|}{~Para~I~}
&\multicolumn{2}{c|}{~Para~II~} \\

\hline

&\multicolumn{2}{c|}{Current}&\multicolumn{1}{c|}{~Future~}&\multicolumn{2}{c|}{Current}&\multicolumn{1}{c|}{~Future~}&\multicolumn{2}{c|}{Current}\\

\hline

&\multicolumn{1}{c|}{~Best~Fit~}&\multicolumn{1}{c|}{~~~~Mean~~~~}&\multicolumn{1}{c|}{~~SD~~}&\multicolumn{1}{c|}{~Best~Fit~}&\multicolumn{1}{c|}{~~~~Mean~~~~}
&\multicolumn{1}{c|}{~~SD~~}&\multicolumn{1}{c|}{~Best~Fit~}&\multicolumn{1}{c|}{~~~~Mean~~~~} \\

\hline

$w_0$&$-1$&$-1$&$-$&$-1.16$&$-1.03^{+0.15+0.36}_{-0.15-0.26}$&$0.045$&$-0.898$&$-0.981^{+0.320+0.534}_{-0.340-0.748}$\\

\hline

$w_1$&$0$&$0$&$-$&$0.968$&$0.405^{+0.562+0.781}_{-0.587-1.570}$&$0.11$&$0.047$&$-0.068^{+0.561+1.037}_{-0.591-1.245}$\\

\hline

$~\Omega_{\DE}~$&$0.760$&$0.762^{+0.015+0.029}_{-0.015-0.033}$&$0.0043$&$0.756$&$0.760^{+0.017+0.033}_{-0.018-0.035}$&$0.0064$&$0.765$&$0.764^{+0.019+0.045}_{-0.019-0.044}$\\

\hline

$H_0$&$73.1$&$73.3^{+1.6+3.2}_{-1.7-3.2}$&$0.44$&$70.3$&$71.2^{+2.3+4.6}_{-2.3-4.2}$&$0.76$&$72.0$&$72.2^{+2.8+5.0}_{-2.6-6.3}$\\

\hline


\end{tabular}
\end{center}
\end{table*}

To study the dynamics of dark energy, we parameterize our universe
as follows:
\begin{equation}
\label{parameterDE} {\bf P} \equiv (\omega_{b}, \omega_{c},
\Theta_{s}, \tau, w_{0}, w_{1}, n_{s}, \log[10^{10}A_{s}])~.
\end{equation}
Our main results of dark energy parameters are summarized in Table
II. Besides the two parameterizations Para I and Para II, we also
investigate the $\Lambda$CDM model for comparison. In addition, we
present the future constraints for the $\Lambda$CDM model and Para
I using the simulated Planck and SNAP data as introduced above.
Marginalized over other cosmological parameters, in Table II we
list the constraints on the dark energy parameters as well as the
Hubble constant in different dark energy models.

\begin{figure}[htbp]
\begin{center}
\includegraphics[scale=0.4]{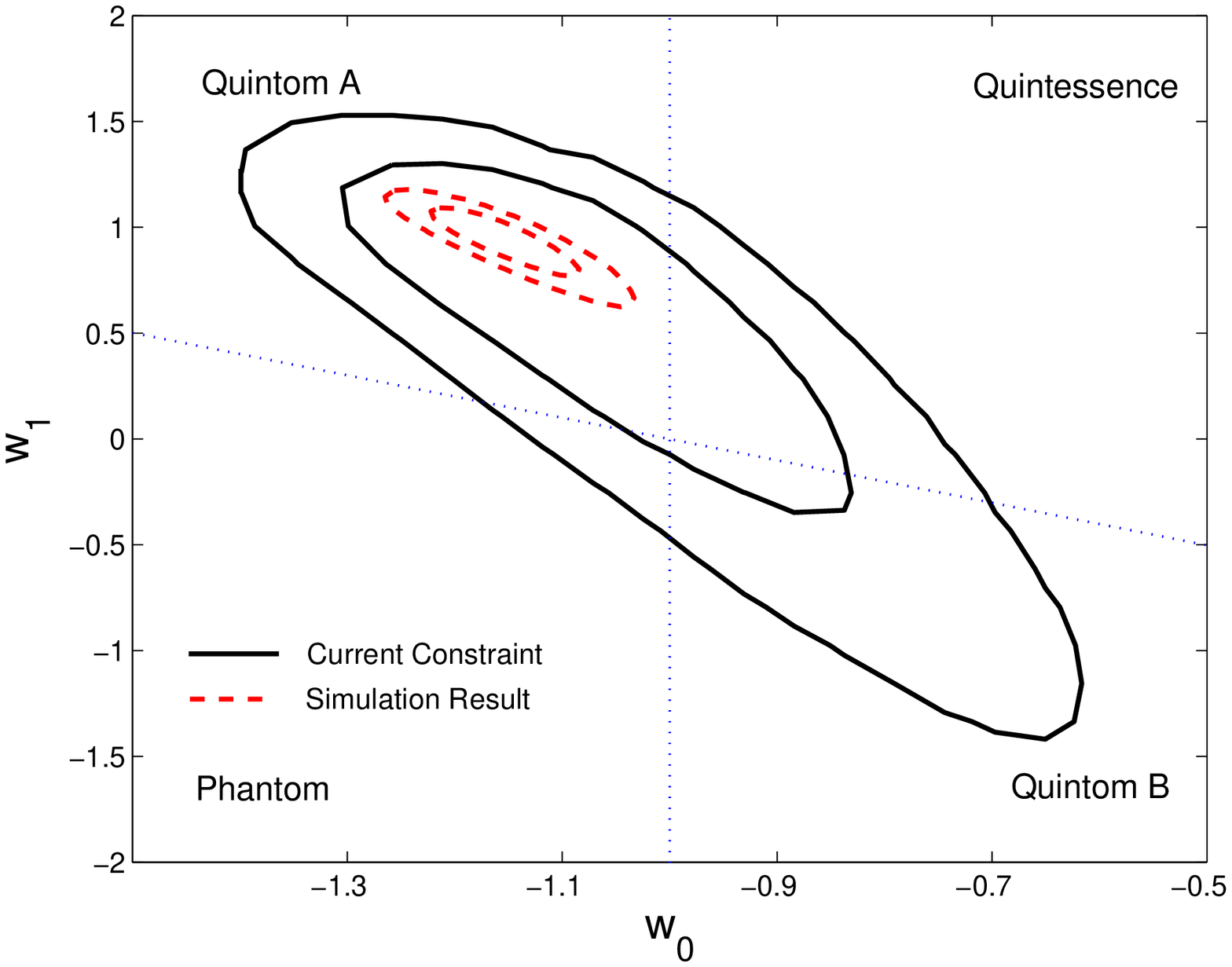}
\includegraphics[scale=0.4]{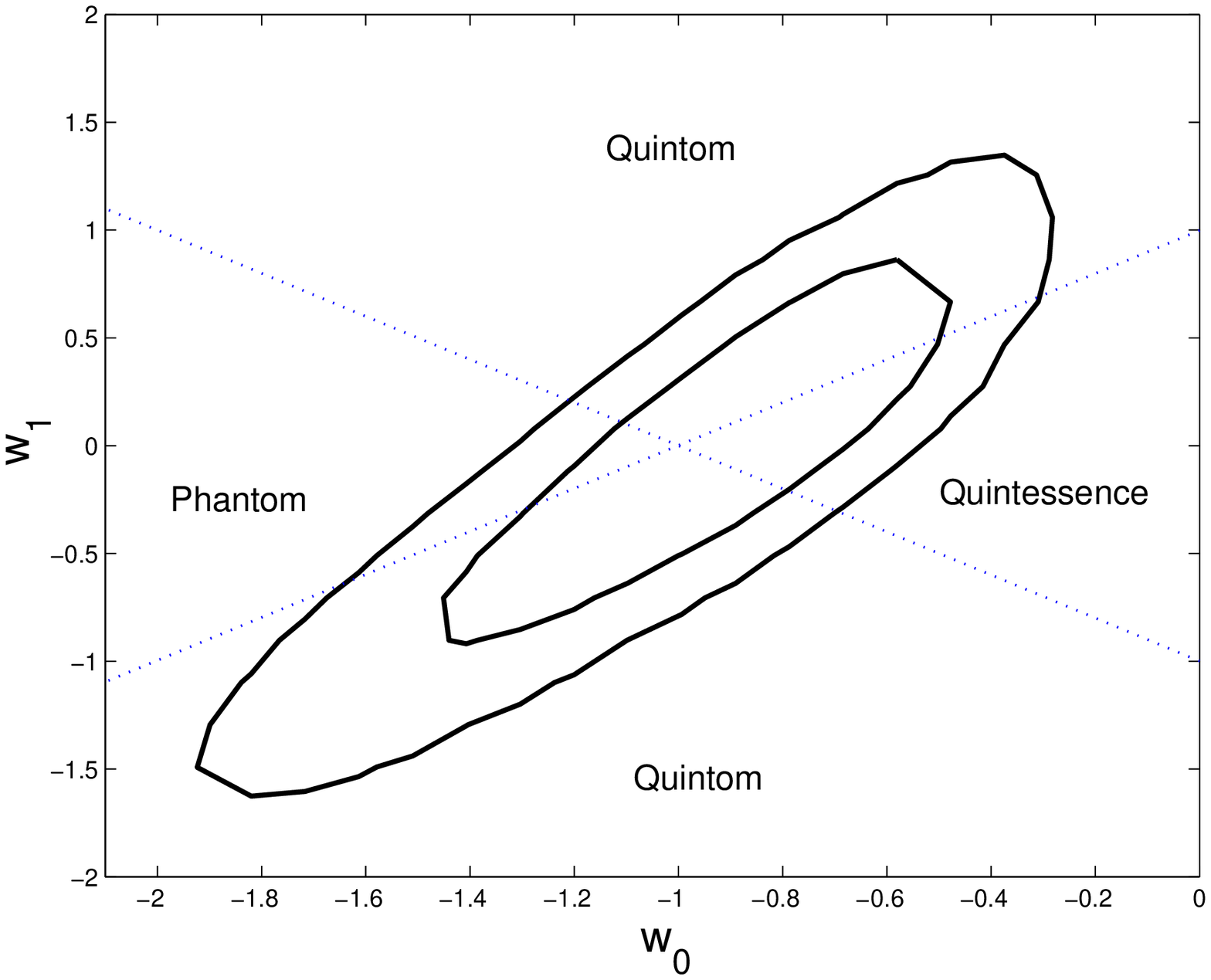}
\caption{Constraints on the dark energy parameters $w_0$ and $w_1$
from the combination of current observations (Black Solid Lines)
and the future simulation data (Red Dashed Lines) respectively.
The left panel is for Para I: $w_{\DE}(a)=w_0+w_1(1-a)$. And the
right panel is for Para II:
$w_{\DE}(a)=w_0+w_1\sin(\frac{3\pi}{2}\log(a))$. The two blue
dotted lines in the ($w_0,w_1$) panel distinguish the dark energy
models and their intersecting point denotes the $\Lambda$CDM
model.\label{fig1}}
\end{center}
\end{figure}

\begin{figure}[htbp]
\begin{center}
\includegraphics[scale=0.4]{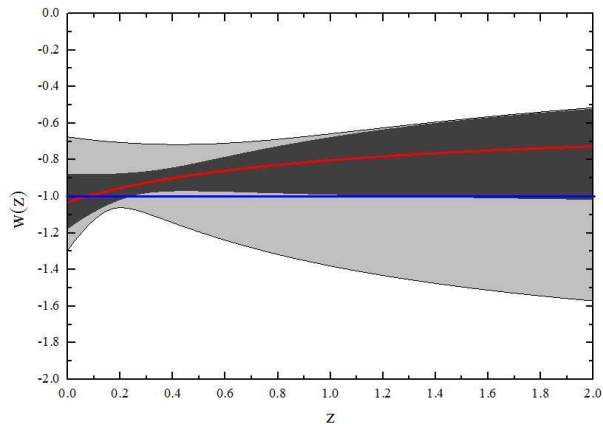}
\caption{Constraints on $w_{\DE}(a)=w_0+w_1(1-a)$ from the current
observations. Median value (central red solid line), $68\%$
(inner, dark shaded area) and $95\%$ (outer, light shaded area)
intervals are shown. The blue dashed lines denote the cosmological
constant boundary.\label{fig2}}
\end{center}
\end{figure}

In Fig.\ref{fig1} we illustrate the constraints on the dark energy
parameters $w_0$ and $w_1$ of two parameterizations. From the
current observations we find that $w_0=-1.03\pm0.15$,
$w_1=0.405^{+0.562}_{-0.587}$ for Para I and the Quintom scenario,
where $w(z)$ can cross the cosmological constant boundary during
the evolution, is mildly favored. Using the current data, we find
that the best fit model is located in the Quintom A region while
the $\Lambda$CDM, denoted by the intersect of two straight lines,
lies at the edge of $1\sigma$ contour. The one dimensional
constraint on the evolution of $w(a)$ from the current data is
shown in Fig.\ref{fig2}. This behavior can be found more obviously
from the best fit model. However, current data can not distinguish
different dark energy models decisively, namely, the variance of
$w_0$ and $w_1$ are too large to distinguish dynamical dark energy
models from the $\Lambda$CDM model. The $\Lambda$CDM model is
still a good fit right now.

In order to distinguish different Dark Energy models we consider
the future measurements Planck and SNAP. The fiducial model we
choose is the best fit model from the current constraints of Para
I. We show the $68\%$ and $95\%$ confidence level contours (Red
Dashed lines) on the left panel of Fig.\ref{fig1}. As we expected,
the constraints from the simulated data are much more stringent
than the current constraints. By the virtue of Planck and SNAP
data, we see that the standard deviations of $w_0$ and $w_1$ are
$\sigma=0.045$ and $\sigma=0.11$ respectively, which are reduced
by a factor of $3.33$ and $5.2$. The Quintom model and the
$\Lambda$CDM model might be distinguished at around $4\sigma$
confidence level.

For Para II, the mean values from the current observations are
$w_0=-0.981^{+0.320}_{-0.340}$, $w_1=-0.068^{+0.561}_{-0.591}$
which still support the Quintom scenario despite of the weak
significance. In right panel of Fig.\ref{fig1}, we see that the
Quintom models occupies the most of the contour while the
$\Lambda$CDM model still lies well within $1\sigma$ contour.


\subsection{Other Cosmological Parameters}

The dynamics of dark energy can have profound effects on the
determination of other cosmological parameters, such as the
inflationary parameters ($n_s$, $\alpha_s$, $r$), the total
neutrino mass $\sum m_{\nu}$ as well as the curvature of
space-time $\Omega_k$, due to the well-known degeneracies among
these parameters. In this subsection, we measure the above
parameters in the framework of dynamical dark energy models.

\begin{table*}
TABLE III. Constraints on some cosmological parameters $n_s$,
$\alpha_s$, $r$, $\Omega_k$ and $\sum m_{\nu}$ from the current
observations and the future simulations. We have shown the mean
$1,2\sigma$ errors (Mean) for the current constraints and the
standard deviation (SD) of these parameters based on the future
simulations. For the weakly constrained parameters we quote the
$95\%$ upper limit instead.
\begin{center}

\begin{tabular}{|c|cc|cc|c|}

\hline

&\multicolumn{2}{c|}{$\Lambda$CDM} &\multicolumn{2}{c|}{~Para~I~}
&\multicolumn{1}{c|}{~Para~II~} \\

\hline

&\multicolumn{1}{c|}{~~~~Mean~~~~}&\multicolumn{1}{c|}{~~SD~~}&\multicolumn{1}{c|}{~~~~Mean~~~~}
&\multicolumn{1}{c|}{~~SD~~}&\multicolumn{1}{c|}{~~~~Mean~~~~} \\

\hline

$n_s$&$0.953^{+0.014+0.028}_{-0.013-0.026}$&$0.003$&$0.965^{+0.017+0.038}_{-0.017-0.032}$&$0.0037$&$0.962^{+0.016+0.036}_{-0.017-0.031}$\\

\hline

$100\times\alpha_s$&$-3.75^{+2.19+4.24}_{-2.21-4.23}$&$0.53$&$-3.38^{+2.52+4.80}_{-2.50-4.76}$&$0.55$&$-3.95^{+2.37+4.86}_{-2.39-4.72}$\\

\hline

$r$&$<0.231$ ($95\%$)&$<0.055$ ($95\%$)&$<0.392$ ($95\%$)&$<0.074$ ($95\%$)&$<0.356$ ($95\%$)\\

\hline

$100\times\Omega_k$&$-0.873^{+0.788+1.454}_{-0.753-1.581}$&$0.289$&$-0.201^{+1.46+2.74}_{-1.29-2.58}$&$1.05$&$-0.593^{+1.23+3.51}_{-1.35-2.57}$\\

\hline

$\sum m_{\nu}$&$<0.958$ ($95\%$)&$0.077$&$<1.59$ ($95\%$)&$0.179$&$<1.53$ ($95\%$)\\

\hline
\end{tabular}
\end{center}
\end{table*}


\subsubsection{Inflationary Models}
\label{Inf}

The current acceleration and the inflation, the two stages of
accelerated expansion of our universe, might have some deep
relationship albeit the significant difference of energy scale
between them. Some efforts have been made to unify these two
expansion epoches, such as the quintessential inflation
\cite{Pvilenkin}. Moreover, the isocurvature perturbations in dark
energy sector generated during inflation may give rise to the
suppression of power of CMB at large scale, which can be mimicked
by suppressed primordial spectrum \cite{Moroi:2003pq}. This means
different dynamics of the dark energy and inflation can lead to
similar effects on observations and studying such degeneracies
might unveil the possible connections between dark energy and
inflation.

\begin{figure}[htbp]
\begin{center}
\includegraphics[scale=0.4]{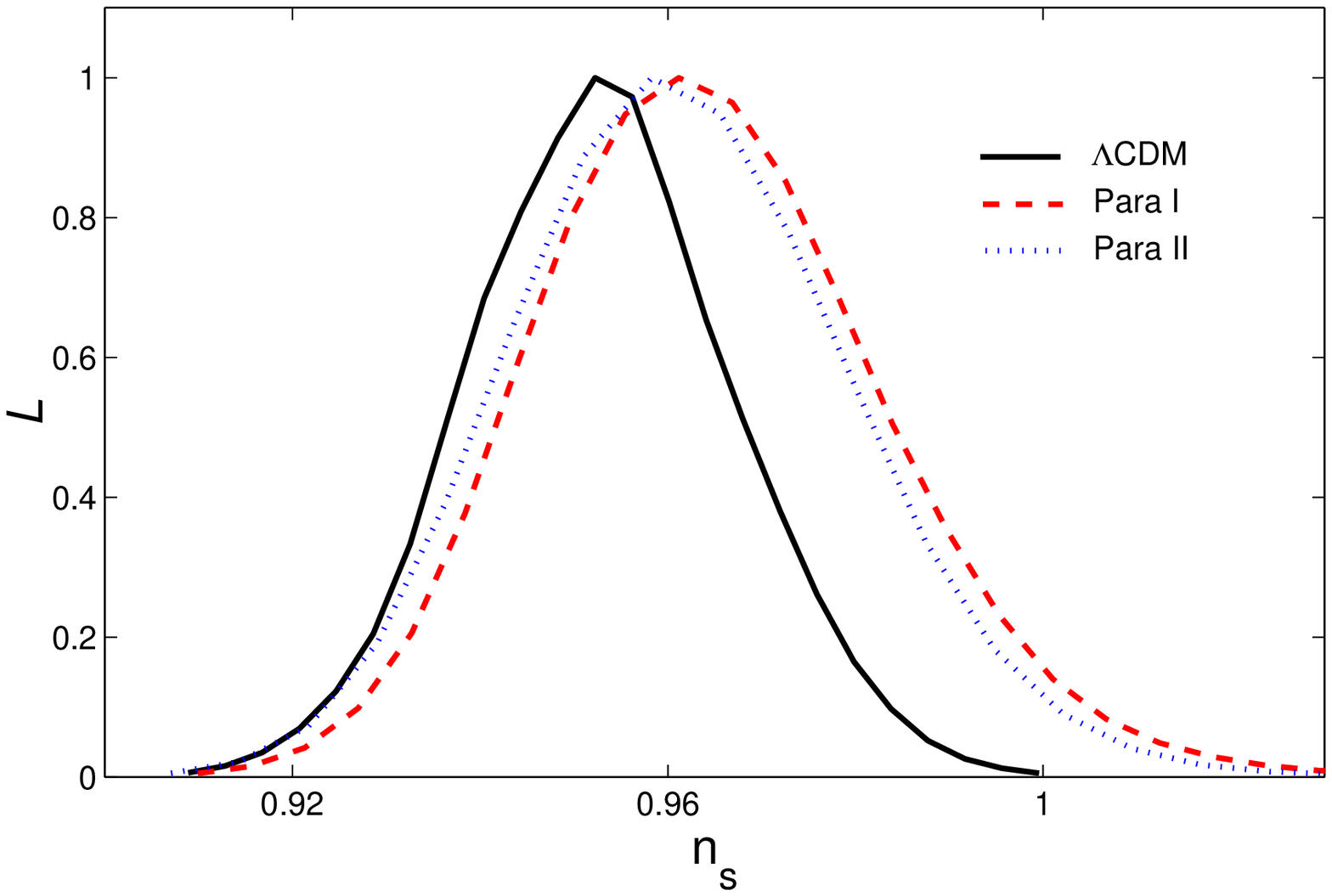}
\includegraphics[scale=0.4]{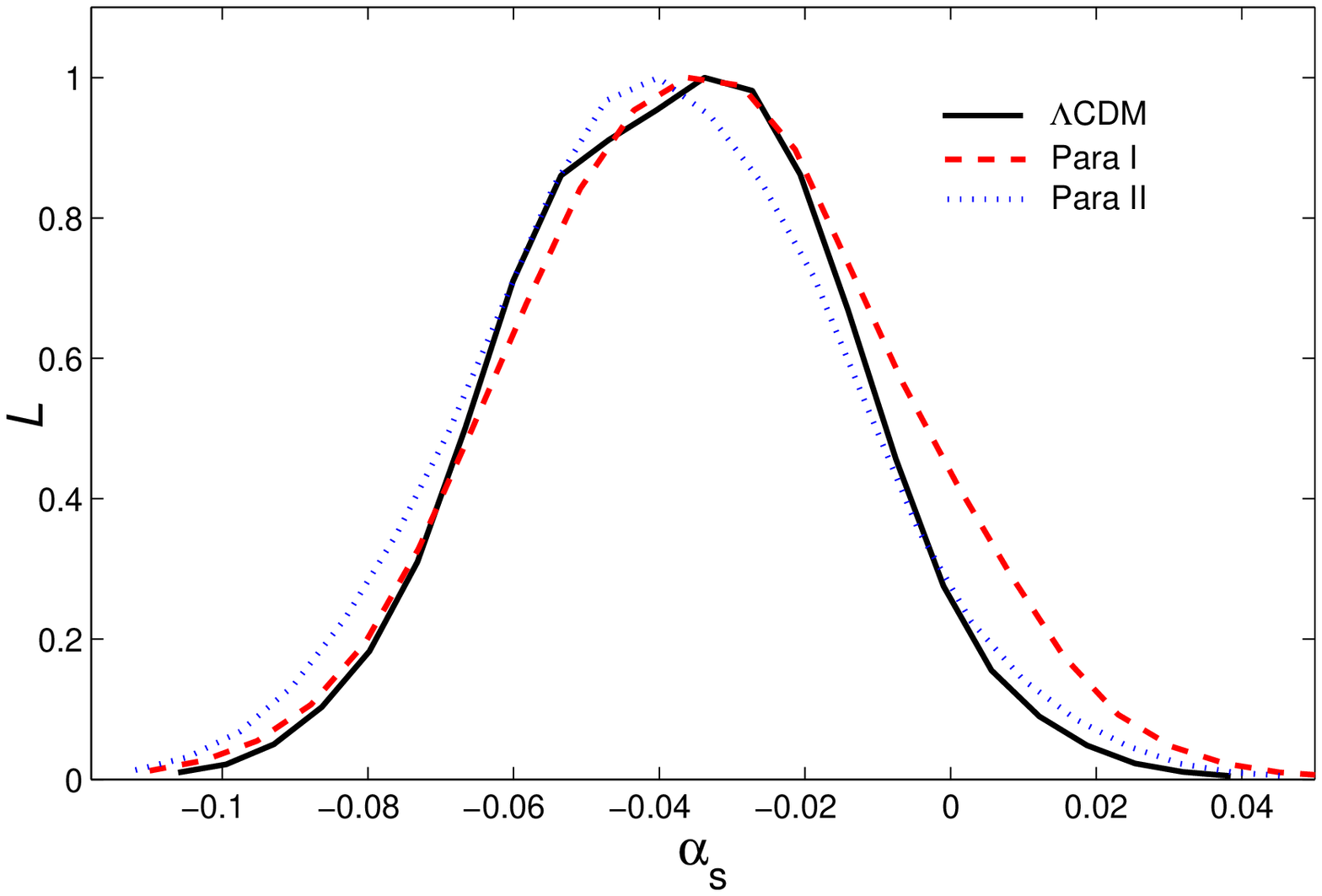}
\includegraphics[scale=0.4]{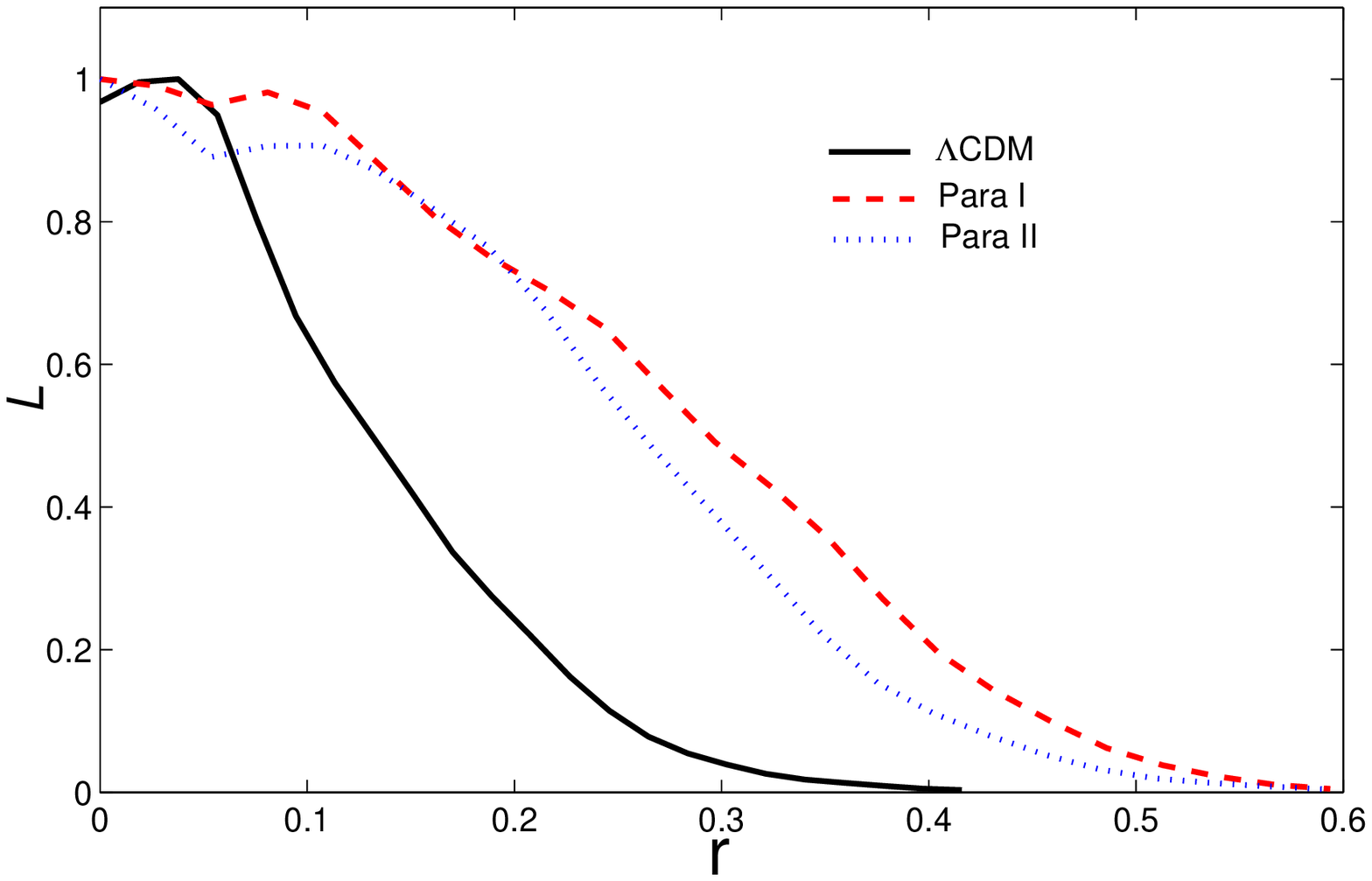}
\caption{$1D$ current constraints on the parameters $n_s$,
$\alpha_s$ and $r$ based on the different dark energy models:
$\Lambda$CDM (black solid line), $w_{\DE}=w_0+w_1(1-a)$ (red
dashed line) and $w_{\DE}=w_0+w_1\sin(\frac{3\pi}{2}\log(a))$
(blue dotted line).\label{fig3}}
\end{center}
\end{figure}

In this section, we constrain the inflationary parameters in the
framework of dynamical dark energy using the current and simulated
datasets. We sample in the following 10 dimensional parameter
space using MCMC algorithm:
\begin{equation}
\label{parameterInf} {\bf P} \equiv (\omega_{b}, \omega_{c},
\Theta_{s}, \tau, w_{0}, w_{1}, n_{s}, \log[10^{10}A_{s}],
\alpha_s, r)~.
\end{equation}
It's noteworthy that we do not constrain $\alpha_s$ and $r$
simultaneously in our global fittings. From Table~III, we can see
the effects of dynamical dark energy on the determination of the
inflationary parameters. Again, we give the fitting results for
Para I, Para II and the $\Lambda$CDM model for comparison. We find
that the constraints for the spectral index $n_s$, the running
$\alpha_s$ and the tensor-to-scalar ratio $r$ have been weaken
with the presence of dynamics of dark energy. Quantitatively, the
$2\sigma$ constraints of $n_s$, $\alpha_s$ and $r$ can be relaxed
by roughly $36\%$, $13\%$ and $70\%$ respectively. This can be
seen from the one dimensional distribution plot of Fig.\ref{fig3}.

The WMAP group found that the scale invariant primordial spectrum
and the inflation models with $n_s>1$ is disfavored at almost the
$3\sigma$ level. Our result is in good agreement with them,
$n_s=0.953^{+0.014}_{-0.013}$, based on the $\Lambda$CDM model.
However, we find that the mean value of $n_s$ moves toward to the
``blue" spectral in the framework of dynamical dark energy model,
$n_s=0.965\pm0.017$. From the future data we find that the
standard deviation of $n_s$ can be shrink to be $\sigma=0.003$ and
the scale invariant spectrum will be verified with much higher
confidence level.

\begin{figure}[htbp]
\begin{center}
\includegraphics[scale=0.4]{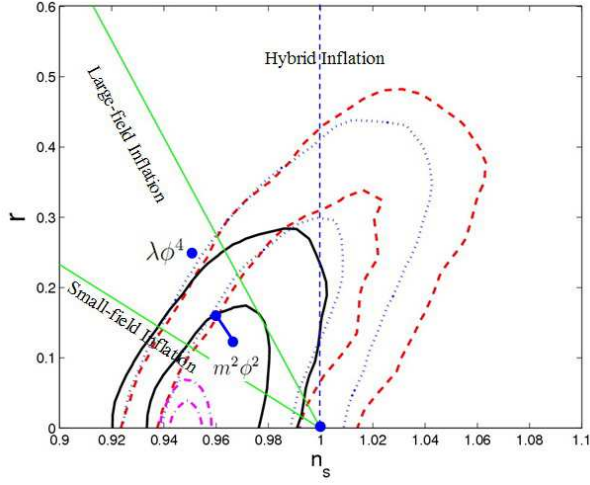}
\caption{$68\%$ and $95\%$ constraints on the panel ($n_s$, $r$)
based on the different Dark Energy models: $\Lambda$CDM (black
solid line), $w_{\DE}=w_0+w_1(1-a)$ (red dashed line) and
$w_{\DE}=w_0+w_1\sin(\frac{3\pi}{2}\log(a))$ (blue dotted line).
The two solid green lines delimit the three classes of inflation
models, namely, small-field, large-field and hybrid models. The
blue points are predicted by $m^2\phi^2$ model and $\lambda\phi^4$
model respectively. These predictions assume that the number of
e-foldings, $N$, is $50-60$ for $m^2\phi^2$ model and $64$ for
$\lambda\phi^4$ model. The magenta dash-dotted lines denote the
$1,2\sigma$ contours obtained from the future simulated
data.\label{fig4}}
\end{center}
\end{figure}

In the framework of dynamical dark energy model, from
Fig.\ref{fig3}, we find that the $95\%$ upper limit of $r$ can be
relax from $r<0.231$ to $r<0.392$. The degeneracy may be due to
the reason that the tensor fluctuation and the dark energy
component mostly affect the large scale (low multipoles) power
spectrum of CMB. In the two dimensional plot of Fig.\ref{fig4}, we
find that the Harrison-Zel'dovich-Peebles scale invariant (HZ)
spectrum ($n_s=1,~r=0$) is disfavored about $2\sim3\sigma$ in
$\Lambda$CDM model. However, this spectrum can be allowed with the
presence of the dynamics of dark energy. The single slow-rolling
scalar field with potential $V(\phi)\sim m^{2}\phi^{2}$, which
predicts $(n_s,r)=(1-2/N,8/N)$, is well within $1\sigma$ region,
while another single slow-rolling scalar field with potential
$V(\phi)\sim \lambda\phi^{4}$, which predicts
$(n_s,r)=(1-3/N,12/N)$, is excluded about $2\sigma$ in the
$\Lambda$CDM model. Interestingly many hybrid inflation models,
excluded in the $\Lambda$CDM model, revive in the framework of
dynamical dark energy model as illustrated in Fig.\ref{fig4}.

Another feature of WMAP data, both for WMAP1 \cite{peiris} and
WMAP3 \cite{wmap3:2006:1,Easther:2006tv}, is the large running of
the scalar primordia spectrum index $\alpha_s$. Our result shows
that the large running is favored more than $1\sigma$,
$\alpha_s=-0.038\pm0.022$. In Fig.\ref{fig3} we find that the
dynamical dark energy models enlarge the error of $\alpha_s$
slightly and do not affect the mean value obviously.

Given the large uncertainties in the constraints of inflationary
parameters from the current observations, different inflation
models cannot be distinguished conclusively. Yet, the constraints
from the future Planck measurement can make this possible. From
our simulation results in Table~III, we find that the error bars
of inflationary parameters can be reduced by about a factor of
$5$. This dramatic improvement will play a crucial role in the
study of dynamics of inflation and can also shed light on the
investigate of the dynamical dark energy model due to the
correlations among inflationary and dark energy parameters.


\subsubsection{Curvature of Universe}
\label{Omk}

Dark energy and the curvature,
$\Omega_K=1-\Omega_{m}-\Omega_{\DE}$, are dominant factors in
determining the fate of our Universe. Further, DE parameters and
$\Omega_K$ are correlated. This is expected since $\Omega_K$ and
dark energy can contribute to the luminosity distance $d_L$ via:
\begin{eqnarray}
\label{lumdis} d_{\rm L}(z)&=&\frac{1+z}{H_0\sqrt{|\Omega_{k}|}}
{\rm sinn}\left[\sqrt{|\Omega_{k}|}\int_0^z
\frac{dz'}{E(z')}\right]~,\\
 E(z)\equiv\frac{H(z)}{H_0}&=&\sqrt{\Omega_m(1+z)^3+\Omega_{\DE}\exp\left(3\int_0^{z}\frac{1+w(z')}{1+z'}dz'\right)
 +\Omega_K(1+z)^2}~,
\end{eqnarray}
where ${\rm sinn}(\sqrt{|k|}x)/\sqrt{|k|}=\sin(x)$, $x$,
$\sinh(x)$ if $k<0$, $k=0$ and $k>0$. In addition, $\Omega_K$ can
modify the angular diameter distance to last scattering surface,
which leaves imprints on the CMB power spectrum.

\begin{figure}[htbp]
\begin{center}
\includegraphics[scale=0.4]{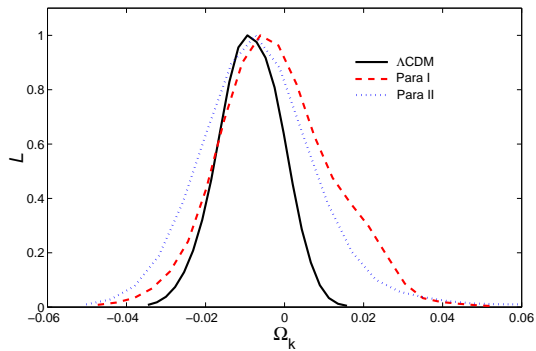}
\caption{$1D$ current constraints on the parameters $\Omega_k$
based on the different dark energy models: $\Lambda$CDM (black
solid line), $w_{\DE}=w_0+w_1(1-a)$ (red dashed line) and
$w_{\DE}=w_0+w_1\sin(\frac{3\pi}{2}\log(a))$ (blue dotted
line).\label{fig5}}
\end{center}
\end{figure}

We concentrate on the determination of $\Omega_K$ in dynamical
dark energy models using current and simulated data. Our parameter
space is:
\begin{equation}
\label{parameter} {\bf P} \equiv (\omega_{b}, \omega_{c},
\Omega_k, \Theta_{s}, \tau, w_{0}, w_{1}, n_{s},
\log[10^{10}A_{s}])~.
\end{equation}
From Table~III and Fig.\ref{fig5}, we see our universe is very
close to flatness, namely, the absolute value of space-time
curvature $|\Omega_k|$ is smaller than $0.025$ in the $\Lambda$CDM
model, $0.028$ for Para I and $0.032$ for Para II. The dynamics of
dark energy weakens the constraint of $|\Omega_k|$ due to the
well-known correlation among $\Omega_k$ and dark energy
parameters. This correlation plays a crucial role in the
reconstruction of equation of state of dark energy
\cite{Clarkson:2007bc}. By the simulated data, we are able to
detect the curvature more accurately.


\subsubsection{Neutrino Mass}
\label{Mnu}

Detecting the absolute mass of neutrino is another challenge of
modern physics. The cosmological observations can obtain upper
limits of the absolute neutrino mass. For background evolution,
neutrino masses, albeit small, contribute to the cosmic energy
budget and modify the epoch of matter-radiation equality, angular
diameter distance to the last scattering surface and other related
physical quantities. For the evolution of perturbation, neutrino
becomes non-relativistic at late time thus they damp the
perturbation within their free streaming scale. Thus the matter
power spectrum can be suppressed by roughly $\Delta P/P \sim - 8
\Omega_\nu/\Omega_m$ \cite{Hu:1997mj}. As a result, neutrino can
leave imprints on the cosmological observations, such as CMB and
matter power spectrum. On the other hand, the evolution of dark
energy can also affect the evolution of background and
perturbation, which mimics the behavior of neutrino to some
extent. This leads to an obvious degeneracy among dark energy
parameters and the neutrino mass.

The degeneracy between dark energy with constant equation of state
and neutrino mass has been studied in the literature
\cite{Hannestad:2005gj,wmap3:2006:1}. In this section, we update
our previous results to study the upper limits of neutrino mass
with the presence of dynamical dark energy \cite{Xia:2006wd} and
investigate the degeneracy between dynamical dark energy and
neutrino mass with current cosmological observations as well as
with the future simulated data.

\begin{figure}[htbp]
\begin{center}
\includegraphics[scale=0.4]{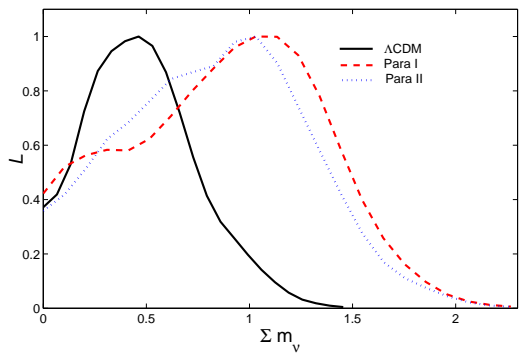}
\caption{$1D$ current constraints on the parameters $\sum m_{\nu}$
based on the different dark energy models: $\Lambda$CDM (black
solid line), $w_{\DE}=w_0+w_1(1-a)$ (red dashed line) and
$w_{\DE}=w_0+w_1\sin(\frac{3\pi}{2}\log(a))$ (blue dotted
line).\label{fig6}}
\end{center}
\end{figure}

We concentrate on the determination of $\sum m_{\nu}$ in dynamical
dark energy models using current and simulated data. Our parameter
space is:
\begin{equation}
\label{parameter} {\bf P} \equiv (\omega_{b}, \omega_{c},
\Theta_{s}, \tau, w_{0}, w_{1}, f_{\nu}, n_{s},
\log[10^{10}A_{s}])~.
\end{equation}
In the last row of Table~III, one can read $95\%~C.L.$ neutrino
mass limits derived from the current observations as well as the
simulated data of Planck and SNAP in the $\Lambda$CDM and
dynamical dark energy models. In the $\Lambda$CDM model, the limit
of neutrino mass we get, $\sum m_{\nu}<0.958~eV~(95\%)$, is
consistent with Tegmark's result \cite{Tegmark:2006az}. For the
dynamical dark energy model, the limit can be relaxed to $\sum
m_{\nu}<1.59~eV~(95\%)$ and $\sum m_{\nu}<1.53~eV~(95\%)$
obviously, due to the degeneracy between the dark energy
parameters and the neutrino mass from the geometric feature of our
Universe \cite{Xia:2006wd,Hannestad:2005gj}. In Fig.\ref{fig6} we
illustrate this effect with current astronomical data.

With the simulated data, we obtain the two tail posterior
distribution due to the nonzero fiducial value of neutrino mass.
The standard deviation will be greatly tightened to be $0.077~eV$.
With the presence of dynamical dark energy, the standard deviation
can be relaxed by a factor of $2.3$ using the simulated data. We
might distinguish the normal hierarchy ($\sum m_{\nu}\sim0.05~eV$)
from the inverted hierarchy ($\sum m_{\nu}\sim0.1~eV$) using the
future Planck measurement.


\subsection{Cosmological \emph{CPT} Violation}
\label{Angle}

The \emph{CPT} symmetry which has been proved to be exact within
the framework of standard model of particle physics and Einstein
gravity could be violated dynamically during the evolution of the
universe \cite{Li:2006ss}. The detection of \emph{CPT} violation
will reveal new physics beyond the standard model. In our previous
work we studied the cosmological \emph{CPT} violation in the
photon sector. We introduce a Chern-Simons term in the effective
Lagrangian of the form \cite{Carroll:1989vb}:
\begin{equation}
\label{CPT}\Delta \mathcal{L} = -\frac{1}{4}p_{\mu}A_{\nu}\tilde
F^{\mu\nu}~,
\end{equation}
where $p_{\mu}$ is a four-vector and $\tilde
F^{\mu\nu}=(1/2)\epsilon^{\mu\nu\rho\sigma}F_{\rho\sigma}$ is the
dual of the electromagnetic tensor. This action is gauge invariant
if $p_{\mu}$ is a constant and homogeneous vector or the gradient
of a scalar. It violates Lorentz and \emph{CPT} symmetries if the
background value of $p_{\mu}$ is nonzero. In the scenario of
quintessential baryo-/leptogenesis \cite{Li:2001st,Li:2002wd} the
four-vector $p_{\mu}$ is in the form of the derivative of the
quintessence scalar, $\partial_{\mu}\phi$. During the evolution of
quintessence, the time component of $\partial_{\mu}\phi$ does not
vanish, which causes \emph{CPT} violation. In the scenario of
gravitational baryo-/leptogenesis
\cite{Davoudiasl:2004gf,Li:2004hh}, $p_{\mu}$ is the gradient of a
function of Ricci scalar R.

\begin{figure}[htbp]
\begin{center}
\includegraphics[scale=0.4]{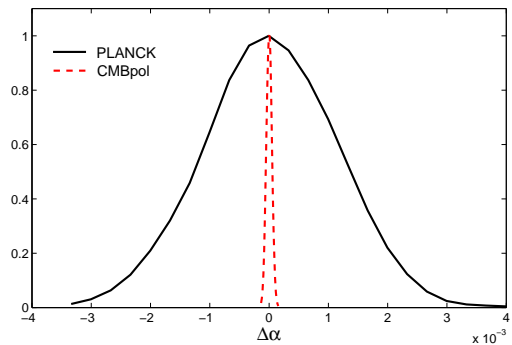}
\caption{$1D$ current constraints on the rotation angle
$\Delta\alpha$ from the simulated Planck (Black Solid line) and
CMBpol (Red Dahsed line) data.\label{fig7}}
\end{center}
\end{figure}

The interaction in Eq.(\ref{CPT}) violates also the \emph{P} and
\emph{CP} symmetries, as long as $p_{0}$ does not vanish
\cite{Klinkhamer:1999zh}. It leads to a rotation of the
polarization vector of electromagnetic waves when they are
propagating over cosmological distances. This effect is known as
``cosmological birefringence". The polarization vector of each
photon is rotated by an angle $\Delta\alpha$ and one would modify
the power spectra of TE, EE, BB, TB and EB at the last scattering
surface as:
\begin{eqnarray}
C_{l}^{'TB} &=& C_{l}^{TE}\sin(2\Delta\alpha)~, \\
C_{l}^{'EB} &=&
\frac{1}{2}(C_{l}^{EE}-C_{l}^{BB})\sin(4\Delta\alpha)~,\\
C_{l}^{'TE} &=& C_{l}^{TE}\cos(2\Delta\alpha)~, \\
C_{l}^{'EE} &=& C_{l}^{EE}\cos^2(2\Delta\alpha) +
C_{l}^{BB}\sin^2(2\Delta\alpha)~,\\
C_{l}^{'BB} &=& C_{l}^{BB}\cos^2(2\Delta\alpha) +
C_{l}^{EE}\sin^2(2\Delta\alpha)~,
\end{eqnarray}
where the primed quantities are rotated. In Ref.\cite{Feng:2006dp}
we have performed a global fit and found that a nonzero rotation
angle of the photons is mildly favored, $\Delta\alpha=-6.0\pm4.0$
deg, using the WMAP3 (without the information of TB and EB power
spectra) and full data of Boomerang-2K2\footnote{The WMAP group
did not release the TB and EB data when we prepared our paper
\cite{Feng:2006dp} at that time. We had to set the TB and EB of
power spectra of WMAP3 data to be zero, $C^{TB}_l=0$ and
$C^{EB}_l=0$. Recently Cabella \emph{et al.} \cite{Cabella:2007br}
performed a wavelet analysis of the temperature and polarization
maps of the CMB delivered by the WMAP experiment which includes
the information of TB and EB power spectra. They obtained a limit
on the CMB photon rotation angle $\Delta\alpha=-2.5\pm3.0$ deg.
Right now the WMAP group has released their results of TB and EB
power spectra. We plan to combine the full data of WMAP3
(including the information of TB and EB power spectra) and
BOOMERANG-2K2 to reanalyze the rotation angle in the future
\cite{xia}.}.

In Fig.\ref{fig7}, by the simulated data with Planck and CMBpol
accuracy, we find that the standard deviation of the rotation
angle will be greatly tightened to $\sigma=0.057$ deg for PLANCK
and $\sigma=2.57\times10^{-3}$ deg for CMBpol. These results are
much more stringent than the current constraint and will be used
to verify the \emph{CPT} symmetry with a higher precision
\cite{Feng:2004mq}.


\section{Summary}
\label{Sum}

Since the mystery of our Universe encodes in the cosmological
parameters, constraining these parameters with the latest
observational data and doing error forecasts with the simulated
futuristic data can lead us to understand the Nature. In this
paper, we focus on the dynamics of dark energy in light of current
and simulated Planck and SNAP data and then constrain the
inflationary parameters, total neutrino mass and curvature of
space-time in the framework of dynamical dark energy model. In
addition, we investigate the rotation angle $\Delta\alpha$, the
possible signature of \emph{CPT} violation, by simulated Planck
and CMBpol data.

Parameterizing the EoS of dark energy in two forms as
Eqs.(\ref{EOS1},\ref{EOS2}), we find that the Quintom model, whose
EoS crosses $-1$ during the evolution, is mildly favored by latest
observations, albeit the $\Lambda$CDM model remains a good fit.
Using the simulated Planck data complimented by SNAP data, we find
that the variance of the dark energy parameters in Eq.(\ref{EOS1})
decreases dramatically, namely, $\sigma(w_0)$ and $\sigma(w_1)$
can be reduced by a factor of $3.33$ and $5.2$ respectively. Given
the current central value, this means that Planck can distinguish
dynamical dark energy from the $\Lambda$CDM model at around
4$\sigma$ confidence level.

Since the dynamics of dark energy greatly affects the
determination of other cosmological parameters, we constrain the
inflationary parameters, total neutrino mass $\sum m_{\nu}$ and
curvature of space-time $\Omega_{K}$ with the presence of dynamics
of dark energy using current and simulated observational data. We
find that the dynamics of dark energy generally weakens the
determination of other cosmological parameters. For instance, we
find that the inflation models with a ``blue" tilt ($n_{s}>1$),
which is strongly disfavored in the $\Lambda$CDM model, are now
well within $2\sigma$ region in the framework of dynamical dark
energy model. This is something intriguing in that Hybrid
Inflation Models have been revived due to the dynamical dark
energy, due to the obviously enlarged parameter space of
($n_s$,$r$) in the framework of dynamical dark energy model. These
discoveries can lead us to further study the dynamics of
inflation, dark energy and the relationship between them. With
Planck, the uncertainties of inflationary parameters, including
$n_{s},~\alpha_{s},~r$, can be roughly reduced by a factor of $5$.
This significant improvement make it possible to distinguish the
inflationary models decidedly.

The presence of dark energy can relax the upper limit of neutrino
mass by a factor of $2$. This sheds new light on the research of
neutrino physics, such as the scenario of mass varying neutrino
\cite{mvn1,mvn2}. By Planck, we can rise the precision of
measurement of this mass limit by a factor of $12$, namely, the
standard deviation of neutrino mass can be shrunk to $0.077~eV$
for the $\Lambda$CDM model and $0.179~eV$ for the dynamical dark
energy model (Para~I). This measurement might make us understand
the Nature of physics conclusively. Current data imply that our
Universe is very close to flatness, say,
$|\Omega_{K}|<0.03~(95\%~C.L.)$. By Planck, we can reduce
$\sigma(\Omega_{K})$ by a factor of $2.67$($\Lambda$CDM). Such
improvement helps us to reveal the Nature of space-time.

The global symmetry of \emph{CPT} plays a critical role in
understanding the fundamental physics. In our previous works we
have found some hints of \emph{CPT} violation encoded in the
rotation angel of polarization vector of photons $\Delta\alpha$.
With Planck and CMBpol, we can test the \emph{CPT} symmetry at a
unprecedent accuracy.


{\bf{Acknowledgments:}} We acknowledge the use of the Legacy
Archive for Microwave Background Data Analysis (LAMBDA). Support
for LAMBDA is provided by the NASA Office of Space Science. We
have performed our numerical analysis on the Shanghai
Supercomputer Center (SSC). We are grateful to Laurence Perotto
for discussions related to the simulation of lensed CMB power
spectra. We thank Yi-Fu Cai, Zu-Hui Fan, Pei-Hong Gu, Steen
Hannestad, Hiranya Peiris, Yun-Song Piao, Levon Pogosian, Tao-Tao
Qiu and Douglas Scott for helpful discussions. This work is
supported in part by National Natural Science Foundation of China
under Grant Nos. 90303004, 10533010 and 10675136 and by the
Chinese Academy of Science under Grant No. KJCX3-SYW-N2.



\begin{thebibliography}{nn}

\bibitem{Miknaitis:2007jd}
  G.~Miknaitis {\it et al.},
  arXiv:astro-ph/0701043.

\bibitem{Davis:2007na}
  T.~M.~Davis {\it et al.},
  arXiv:astro-ph/0701510.

\bibitem{wmap3:2006:1}
  D.~N.~Spergel {\it et al.},
  Astrophys.\ J.\ Suppl.\  {\bf 170}, 377 (2007).


\bibitem{wmap3:2006:2}
  L.~Page {\it et al.},
  Astrophys.\ J.\ Suppl.\  {\bf 170}, 335 (2007).


\bibitem{wmap3:2006:3}
  G.~Hinshaw {\it et al.},
  Astrophys.\ J.\ Suppl.\  {\bf 170}, 288 (2007).


\bibitem{wmap3:2006:4}
  N.~Jarosik {\it et al.},
  Astrophys.\ J.\ Suppl.\  {\bf 170}, 263 (2007).


\bibitem{Tegmark:2006az}
  M.~Tegmark {\it et al.},
  Phys.\ Rev.\  D {\bf 74}, 123507 (2006).

\bibitem{Feng:2006dp}
  B.~Feng, M.~Li, J.~Q.~Xia, X.~Chen and X.~Zhang,
  Phys.\ Rev.\ Lett.\  {\bf 96}, 221302 (2006).

\bibitem{SN98}
A. G. Riess {\it et al.}, Astron. J. {\bf 116}, 1009 (1998); S.
Perlmutter {\it et al.}, Astrophys. J. {\bf 517}, 565 (1999).

\bibitem{SW89}
S. Weinberg, Rev. Mod. Phys. {\bf 61}, 1 (1989); I. Zlatev, L.-M.
Wang, and P. J. Steinhardt, Phys. Rev. Lett. {\bf 82}, 896 (1999).

\bibitem{quint}
B. Ratra and P. J. E. Peebles, Phys. Rev. D {\bf 37}, 3406 (1988);
P. J. E. Peebles and B. Ratra, Astrophys. J. {\bf 325}, L17 (1988);
C. Wetterich, Nucl. Phys. B {\bf 302}, 668 (1988); C. Wetterich,
Astron. Astrophys. {\bf 301}, 321 (1995).

\bibitem{phantom}
R. R. Caldwell, Phys. Lett. B {\bf 545}, 23 (2002).

\bibitem{kessence}
C. Armendariz-Picon, V. Mukhanov and P. J. Steinhardt, Phys. Rev.
Lett. {\bf 85}, 4438 (2000);~Phys. Rev. D {\bf 63}, 103510 (2001).


\bibitem{quintom}
  B.~Feng, X.~L.~Wang and X.~M.~Zhang,
  Phys.\ Lett.\ B {\bf 607}, 35 (2005).

\bibitem{Xia:2005ge}
  J.~Q.~Xia, G.~B.~Zhao, B.~Feng, H.~Li and X.~Zhang,
  Phys.\ Rev.\  D {\bf 73}, 063521 (2006).

\bibitem{Xia:2006cr}
  J.~Q.~Xia, G.~B.~Zhao, B.~Feng and X.~Zhang,
  JCAP {\bf 0609}, 015 (2006).

\bibitem{Zhao:2006bt}
  G.~B.~Zhao, J.~Q.~Xia, B.~Feng and X.~Zhang,
  Int.\ J.\ Mod.\ Phys.\  D {\bf 16}, 1229 (2007).


\bibitem{Xia:2006wd}
  J.~Q.~Xia, G.~B.~Zhao and X.~Zhang,
  Phys.\ Rev.\  D {\bf 75}, 103505 (2007).

\bibitem{Zhao:2006qg}
  G.~B.~Zhao, J.~Q.~Xia, H.~Li, C.~Tao, J.~M.~Virey, Z.~H.~Zhu and X.~Zhang,
  Phys.\ Lett.\  B {\bf 648}, 8 (2007).

\bibitem{others}
  D.~Huterer and A.~Cooray,
  Phys.\ Rev.\ D {\bf 71}, 023506 (2005);
  U.~Alam, V.~Sahni and A.~A.~Starobinsky,
  JCAP {\bf 0702}, 011 (2007);
  V.~Barger, Y.~Gao and D.~Marfatia,
  Phys.\ Lett.\  B {\bf 648}, 127 (2007);
  H.~Li, M.~Su, Z.~Fan, Z.~Dai and X.~Zhang,
  Phys.\ Lett.\  B {\bf 658}, 95 (2008).

\bibitem{Xia:2007km}
  J.~Q.~Xia, Y.~F.~Cai, T.~T.~Qiu, G.~B.~Zhao and X.~Zhang,
  arXiv:astro-ph/0703202.

\bibitem{Vikman:2004dc}
  A.~Vikman,
  Phys.\ Rev.\  D {\bf 71}, 023515 (2005).

\bibitem{Kunz:2006wc}
  M.~Kunz and D.~Sapone,
  Phys.\ Rev.\  D {\bf 74}, 123503 (2006).

\bibitem{study4quintom}
H. Wei and R. G. Cai, Class.\ Quant.\ Grav.\  {\bf 22}, 3189
(2005); R. G. Cai, H. S. Zhang and A. Wang, Commun.\ Theor.\
Phys.\  {\bf 44}, 948 (2005); A. A. Andrianov, F. Cannata and A.
Y. Kamenshchik, Phys.\ Rev.\  D {\bf 72}, 043531 (2005); X. Zhang,
Int.\ J.\ Mod.\ Phys.\  D {\bf 14}, 1597 (2005); Q. Guo and R. G.
Cai, gr-qc/0504033; B. McInnes, Nucl. Phys. B {\bf 718}, 55
(2005); E. Elizalde, S. Nojiri, S. D. Odintsov and P. Wang, Phys.
Rev. D {\bf 71}, 103504 (2005); I. Y. Aref'eva, A. S. Koshelev,
and S. Yu. Vernov, Phys.\ Rev.\  D {\bf 72}, 064017 (2005); A.
Anisimov, E. Babichev and A. Vikman, J. Cosmol. Astropart. Phys.
{\bf 0506}, 006 (2005); H. Stefancic, Phys.\ Rev.\  D {\bf 71},
124036 (2005); J.~Zhang, X.~Zhang and H.~Liu,
arXiv:astro-ph/0612642.


\bibitem{inflation}
  A.~H.~Guth,
  Phys.\ Rev.\ D {\bf 23}, 347 (1981);
  K.~Sato,
  Mon.\ Not.\ Roy.\ Astron.\ Soc.\  {\bf 195}, 467 (1981);
For a
  review see e.g.
  A.~D.~Linde,
  Phys.\ Rept.\  {\bf 333}, 575 (2000).

\bibitem{Guth:1979bh}
For relevant studies see also
  A.~H.~Guth and S.~H.~H.~Tye,
  Phys.\ Rev.\ Lett.\  {\bf 44}, 631 (1980)
  [Erratum-ibid.\  {\bf 44}, 963 (1980)];
  A.~A.~Starobinsky,
  Phys.\ Lett.\ B {\bf 91}, 99 (1980).

\bibitem{Easther:2006tv}
  R.~Easther and H.~Peiris,
  JCAP {\bf 0609}, 010 (2006).

\bibitem{:2006uk}
  Planck Collaboration,
  arXiv:astro-ph/0604069.

\bibitem{Zhao:2005vj}
  G.~B.~Zhao, J.~Q.~Xia, M.~Li, B.~Feng and X.~Zhang,
  Phys.\ Rev.\  D {\bf 72}, 123515 (2005).

\bibitem{CosmoMC}
  A.~Lewis and S.~Bridle,
  Phys.\ Rev.\ D {\bf 66}, 103511 (2002).

\bibitem{paraPk}
  A.~Kosowsky and M.~S.~Turner,
  Phys.\ Rev.\ D {\bf 52}, 1739 (1995);
  J.~E.~Lidsey, A.~R.~Liddle, E.~W.~Kolb, E.~J.~Copeland, T.~Barreiro and M.~Abney,
  Rev.\ Mod.\ Phys.\  {\bf 69}, 373 (1997);
  S.~Hannestad, S.~H.~Hansen, F.~L.~Villante and A.~J.~S.~Hamilton,
  Astropart.\ Phys.\  {\bf 17}, 375 (2002);
  S.~L.~Bridle, A.~M.~Lewis, J.~Weller and G.~Efstathiou,
  Mon.\ Not.\ Roy.\ Astron.\ Soc.\  {\bf 342}, L72 (2003);
  B.~Feng, X.~Gong and X.~Wang,
Mod.\ Phys.\ Lett.\ A {\bf
19}, 2377 (2004). 

\bibitem{Linderpara}
  M.~Chevallier and D.~Polarski,
  Int.\ J.\ Mod.\ Phys.\ D {\bf 10}, 213 (2001).

\bibitem{Xia:2006rr}
  J.~Q.~Xia, G.~B.~Zhao, H.~Li, B.~Feng and X.~Zhang,
  Phys.\ Rev.\  D {\bf 74}, 083521 (2006).

\bibitem{Feng:2004ff}
  B.~Feng, M.~Li, Y.~S.~Piao and X.~Zhang,
  Phys.\ Lett.\  B {\bf 634}, 101 (2006).

\bibitem{Barenboim:2004kz}
  G.~Barenboim, O.~Mena and C.~Quigg,
  Phys.\ Rev.\  D {\bf 71}, 063533 (2005).

\bibitem{Xia:2004rw}
  J.~Q.~Xia, B.~Feng and X.~M.~Zhang,
  Mod.\ Phys.\ Lett.\  A {\bf 20}, 2409 (2005).

\bibitem{Riess:2006fw}
  A.~G.~Riess {\it et al.},
  arXiv:astro-ph/0611572.

\bibitem{ma}
  C.~P.~Ma and E.~Bertschinger,
  Astrophys.\ J.\  {\bf 455}, 7 (1995).


\bibitem{MacTavish:2005yk}
  C.~J.~MacTavish {\it et al.},
  Astrophys.\ J.\  {\bf 647}, 799 (2006).

\bibitem{Readhead:2004gy}
  A.~C.~S.~Readhead {\it et al.},
  Astrophys.\ J.\  {\bf 609}, 498 (2004).

\bibitem{Dickinson:2004yr}
  C.~Dickinson {\it et al.},
  Mon.\ Not.\ Roy.\ Astron.\ Soc.\  {\bf 353}, 732 (2004).

\bibitem{Kuo:2002ua}
  C.~l.~Kuo {\it et al.},
  Astrophys.\ J.\  {\bf 600}, 32 (2004).



\bibitem{Cole:2005sx}
  S.~Cole {\it et al.},
  Mon.\ Not.\ Roy.\ Astron.\ Soc.\  {\bf 362} (2005) 505.

\bibitem{DiPietro:2002cz}
  E.~Di Pietro and J.~F.~Claeskens,
  Mon.\ Not.\ Roy.\ Astron.\ Soc.\  {\bf 341}, 1299 (2003).


\bibitem{Hubble}
  W.~L.~Freedman {\it et al.},
  Astrophys.\ J.\  {\bf 553}, 47 (2001).

\bibitem{BBN}
  S.~Burles, K.~M.~Nollett and M.~S.~Turner,
  Astrophys.\ J.\  {\bf 552}, L1 (2001).


\bibitem{R-1}
  A. Gelman and D. Rubin,
  Statistical Science {\bf 7}, 457 (1992).



\bibitem{Easther:2004vq}
  R.~Easther, W.~H.~Kinney and H.~Peiris,
  JCAP {\bf 0505}, 009 (2005).

\bibitem{Perotto:2006rj}
  L.~Perotto, J.~Lesgourgues, S.~Hannestad, H.~Tu and Y.~Y.~Y.~Wong,
  JCAP {\bf 0610}, 013 (2006).



\bibitem{Zaldarriaga:1998ar}
  M.~Zaldarriaga and U.~Seljak,
  Phys.\ Rev.\  D {\bf 58}, 023003 (1998).

\bibitem{Lewis:2006fu}
  A.~Lewis and A.~Challinor,
  Phys.\ Rept.\  {\bf 429}, 1 (2006).

\bibitem{Lewis:1999bs}
  A.~Lewis, A.~Challinor and A.~Lasenby,
  Astrophys.\ J.\  {\bf 538}, 473 (2000).

\bibitem{Hu:2001fb}
  W.~Hu,
  Phys.\ Rev.\  D {\bf 65}, 023003 (2002).

\bibitem{Smith:2006nk}
  K.~M.~Smith, W.~Hu and M.~Kaplinghat,
  Phys.\ Rev.\  D {\bf 74}, 123002 (2006).

\bibitem{Hu:2001kj}
  W.~Hu and T.~Okamoto,
  Astrophys.\ J.\  {\bf 574}, 566 (2002).

\bibitem{kim} A.~G.~Kim, E.~V.~Linder, R.~Miquel and N.~Mostek,
  Mon.\ Not.\ Roy.\ Astron.\ Soc.\  {\bf 347}, 909 (2004).

\bibitem{Li:2005zd}
  H.~Li, B.~Feng, J.~Q.~Xia and X.~Zhang,
   Phys.\ Rev.\ D {\bf 73}, 103503 (2006).

\bibitem{Pvilenkin}
  P.~J.~E.~Peebles and A.~Vilenkin,
  Phys.\ Rev.\ D {\bf 59}, 063505 (1999).

\bibitem{Moroi:2003pq}
  T.~Moroi and T.~Takahashi,
  Phys.\ Rev.\ Lett.\  {\bf 92}, 091301 (2004).

\bibitem{peiris}
  H.~V.~Peiris {\it et al.},
  Astrophys.\ J.\ Suppl.\  {\bf 148}, 213 (2003).

\bibitem{Clarkson:2007bc}
  C.~Clarkson, M.~Cortes and B.~A.~Bassett,
  JCAP {\bf 0708}, 011 (2007).

\bibitem{Hu:1997mj}
  W.~Hu, D.~J.~Eisenstein and M.~Tegmark,
  Phys.\ Rev.\ Lett.\  {\bf 80}, 5255 (1998).

\bibitem{Hannestad:2005gj}
  S.~Hannestad,
  Phys.\ Rev.\ Lett.\  {\bf 95}, 221301 (2005).

\bibitem{Li:2006ss}
  M.~Li, J.~Q.~Xia, H.~Li and X.~Zhang,
  Phys.\ Lett.\  B {\bf 651}, 357 (2007).


\bibitem{Carroll:1989vb}
  S.~M.~Carroll, G.~B.~Field and R.~Jackiw,
  Phys.\ Rev.\  D {\bf 41}, 1231 (1990).

\bibitem{Li:2001st}
  M.~z.~Li, X.~l.~Wang, B.~Feng and X.~m.~Zhang,
  Phys.\ Rev.\  D {\bf 65}, 103511 (2002).

\bibitem{Li:2002wd}
  M.~Li and X.~Zhang,
  Phys.\ Lett.\  B {\bf 573}, 20 (2003).


\bibitem{Davoudiasl:2004gf}
  H.~Davoudiasl, R.~Kitano, G.~D.~Kribs, H.~Murayama and P.~J.~Steinhardt,
  Phys.\ Rev.\ Lett.\  {\bf 93}, 201301 (2004).

\bibitem{Li:2004hh}
  H.~Li, M.~z.~Li and X.~m.~Zhang,
  Phys.\ Rev.\  D {\bf 70}, 047302 (2004).

\bibitem{Klinkhamer:1999zh}
  F.~R.~Klinkhamer,
  Nucl.\ Phys.\  B {\bf 578}, 277 (2000).


\bibitem{Cabella:2007br}
  P.~Cabella, P.~Natoli and J.~Silk,
  arXiv:0705.0810 [astro-ph].

\bibitem{xia}
  J.~Q.~Xia, H.~Li, X.~l.~Wang and X.~m.~Zhang,
  arXiv:0710.3325 [hep-ph].

\bibitem{Feng:2004mq}
  B.~Feng, H.~Li, M.~z.~Li and X.~m.~Zhang,
  Phys.\ Lett.\  B {\bf 620}, 27 (2005).

\bibitem{mvn1}
  P.~Gu, X.~Wang and X.~Zhang,
  Phys.\ Rev.\  D {\bf 68}, 087301 (2003);
  R.~Fardon, A.~E.~Nelson and N.~Weiner,
  JCAP {\bf 0410}, 005 (2004);
  R.~D.~Peccei,
  Phys.\ Rev.\  D {\bf 71}, 023527 (2005).

\bibitem{mvn2}
  G.~B.~Zhao, J.~Q.~Xia and X.~M.~Zhang,
  JCAP {\bf 0707}, 010 (2007).

\end{thebibliography}
\end{document}